\documentclass[11pt]{article}
\usepackage{jheppub}
\usepackage{bm,bbm}
\usepackage{booktabs,amsmath}
\usepackage{mathtools}
\usepackage{graphics}
\usepackage{braket}
\usepackage{tikz}
\usepackage{ascmac}
\usepackage{enumitem}
\usepackage{comment}
\usepackage{stackrel}
\usepackage{accents}
\usepackage{cases}
\usepackage{youngtab} 
\usepackage{extarrows}
\usepackage{cancel}
\usepackage{physics}
\usepackage{appendix}
\usetikzlibrary{shapes}
\usepackage{framed}
\usepackage{colortbl}
\usepackage[normalem]{ulem}

\usepackage{xcolor}

\definecolor{shadecolor}{rgb}{0.90,0.90,0.90}
\definecolor{color1}{rgb}{0.94,0.86,0.73}
\definecolor{color2}{rgb}{0.94,0.23,0.38}
\definecolor{color3}{rgb}{0.0,0.43,0.72}
\definecolor{color1a}{rgb}{0.55,0.62,0.38}
\definecolor{color2a}{rgb}{0.91,0.60,0.61}
\definecolor{color3a}{rgb}{0.52,0.65,0.73}
\definecolor{purple2}{rgb}{0.8,0.0,0.8}
\usetikzlibrary{patterns}
\usetikzlibrary{decorations.markings}
\usetikzlibrary{decorations.pathmorphing}
\tikzset{snake it/.style={decorate, decoration=snake}}

\makeatletter
\newcommand{\mylabel}[2]{#2\def\@currentlabel{#2}\label{#1}}
\makeatother

\usetikzlibrary{quotes,angles}

\usetikzlibrary{shadings}

\tikzset{test/.style n args={3}{
    postaction={
    decorate,
    decoration={
    markings,
    mark=between positions 0 and \pgfdecoratedpathlength step 0.5pt with {
    \pgfmathsetmacro\myval{multiply(
        divide(
        \pgfkeysvalueof{/pgf/decoration/mark info/distance from start}, \pgfdecoratedpathlength
        ),
        100
    )};
    \pgfsetfillcolor{#3!\myval!#2};
    \pgfpathcircle{\pgfpointorigin}{#1};
    \pgfusepath{fill};}
}}}}

\definecolor{c1}{rgb}{0.0,0.2,0.9}
\definecolor{c2}{rgb}{0.45,0.0,0.45}
\definecolor{c3}{rgb}{0.9,0.2,0.0}

\usetikzlibrary{arrows}
\usetikzlibrary {arrows.meta}

\tikzset{middlearrow/.style={
        decoration={markings,
            mark= at position 0.57 with {\arrow{#1}} ,
        },
        postaction={decorate}
    }
}

\usepackage{blkarray}

\usepackage{nicematrix}
\NiceMatrixOptions%
{code-for-first-row = \scriptstyle,
code-for-first-col = \scriptstyle }

\usetikzlibrary{decorations.markings}
\def\MarkLt{4pt}
\def\MarkSep{2pt}

\tikzset{
  TwoMarks/.style={
    postaction={decorate,
      decoration={
        markings,
        mark=at position #1 with
          {
              \begin{scope}[xslant=0.2]
              \draw[line width=\MarkSep,white,-] (0pt,-\MarkLt) -- (0pt,\MarkLt) ;
              \draw[-] (-0.5*\MarkSep,-\MarkLt) -- (-0.5*\MarkSep,\MarkLt) ;
              \draw[-] (0.5*\MarkSep,-\MarkLt) -- (0.5*\MarkSep,\MarkLt) ;
              \end{scope}
          }
       }
    }
  },
  TwoMarks/.default={0.5},
  OneMark/.style={
    postaction={decorate,
      decoration={
        markings,
        mark=at position #1 with
          {
              \draw[-] (0,-\MarkLt) -- (0,\MarkLt) ;
          }
       }
    }
  },
  OneMark/.default={0.5}
}

\usepackage{tikz}
\usetikzlibrary{intersections, calc,positioning,decorations.pathreplacing,decorations.pathmorphing}
\tikzset{>=latex}

\def\i{{\rm i}}

\def\b0{\bm{0}_\perp}

\usepackage{centernot}
\usepackage{mathtools}
\usepackage{stmaryrd}

\makeatletter
\newcommand{\xMapsto}[2][]{\ext@arrow 0599{\Mapstofill@}{#1}{#2}}
\def\Mapstofill@{\arrowfill@{\Mapstochar\Relbar}\Relbar\Rightarrow}
\makeatother

\usepackage{multirow}

\usepackage{tikz-3dplot}

\DeclareFontFamily{U}{mathx}{\hyphenchar\font45}
\DeclareFontShape{U}{mathx}{m}{n}{
      <5> <6> <7> <8> <9> <10>
      <10.95> <12> <14.4> <17.28> <20.74> <24.88>
      mathx10
      }{}
\DeclareSymbolFont{mathx}{U}{mathx}{m}{n}
\DeclareFontSubstitution{U}{mathx}{m}{n}
\DeclareMathAccent{\widecheck}{0}{mathx}{"71}
\newcommand{\bea}{\begin{eqnarray}}
\newcommand{\eea}{\end{eqnarray}}

\title{Generalizations and UV completions of Cho--Maison monopole}

\author{Fukutaro Miya}
\emailAdd{fkmiya@het.phys.sci.osaka-u.ac.jp}

\author{and Ryosuke Sato}
\emailAdd{rsato@het.phys.sci.osaka-u.ac.jp}

\affiliation{Department of Physics, The University of Osaka, Toyonaka, Osaka 560-0043, Japan}

\abstract{A monopole configuration in the electroweak theory was constructed by Cho and Maison~\cite{Cho:1996qd}, allowing for a singular behavior at the origin. 
Since the essential structure of the Cho--Maison monopole is based on an electroweak-type symmetry breaking, similar monopole configurations are expected to arise more generally in gauge theories containing such a structure.
In this paper, we explicitly show that Cho--Maison-like monopole configurations can indeed be constructed in a broad class of models. 
We also show that the Cho--Maison monopole can be embedded into an 't~Hooft--Polyakov monopole as its low-energy effective description. 
In particular, we find that a monopole in the Pati--Salam model behaves as the electroweak Cho--Maison monopole once degrees of freedom which are heavier than the electroweak scale are integrated out.}
\begin{document}
\begin{flushright}
OU-HET-1314
\end{flushright}
\maketitle
\flushbottom

\section{Introduction}
Magnetic monopoles have long been a subject of fundamental interest in theoretical physics. 
The Dirac monopole \cite{Dirac:1948um} provides the first example of such an object, demonstrating that the mere existence of a single monopole explains the quantization of electric charge. 
A nonsingular, finite-energy monopole solution, known as the ’t~Hooft--Polyakov monopole, was first constructed independently by ’t~Hooft \cite{tHooft:1974kcl} and Polyakov \cite{Polyakov:1974ek} in an $\mathrm{SU(2)}$ gauge theory spontaneously broken to $\mathrm{U(1)}$. 
In a model with gauge symmetry breaking $G \to H$, the set of vacua forms a manifold $G/H$,
and the 't Hooft--Polyakov monopoles exist if $\pi_2(G/H)$ is non-trivial. 
One of the most important applications of this argument is grand unified theories (GUTs). 
In particular, in typical GUT scenarios such as 
\(\mathrm{SU(5)}\) \cite{Georgi:1974sy} and the Pati--Salam model \cite{Pati:1974yy}, both of which are broken down to the Standard Model gauge group \(\mathrm{SU(3)_C \times SU(2)_L \times U(1)_Y}\), one finds \(\pi_2(G/H)\simeq \mathbb{Z}\), 
implying the existence of topologically stable ’t~Hooft--Polyakov monopoles.

In contrast, the Standard Model does not admit an ’t~Hooft--Polyakov monopole.
This is because the electroweak symmetry breaking
\(\mathrm{SU(2)_{L}}\times \mathrm{U(1)_{Y}} \to \mathrm{U(1)_{\mathrm{EM}}}\)
results in the vacuum manifold
\(\mathcal{M}_{\rm SM} \simeq (\mathrm{SU(2)_{L}}\times \mathrm{U(1)_{Y}})/\mathrm{U(1)_{\mathrm{EM}}} \simeq S^3\), for which \(\pi_{2}(S^{3})=0\).
This, however, does not exclude the existence of monopole-like configurations. 
Indeed, Cho and Maison \cite{Cho:1996qd} showed that an electroweak monopole solution can be constructed.  
In analogy with the Wu--Yang approach \cite{Wu:1976ge}, the configuration is defined by gluing two gauge patches, thereby avoiding an explicit Dirac string singularity.
Although the Cho--Maison monopole was found 30 years ago, as we will demonstrate in this paper, several aspects of this configuration remain to be clarified.

First, the Cho--Maison monopole has been discussed only in models with electroweak gauge symmetry breaking, to the best of our knowledge. 
However, it is important to investigate under what conditions the Cho--Maison monopole configurations can arise, and how their existence is related to the structure of symmetry breaking. 
This situation is in contrast to the ’t~Hooft--Polyakov monopole, whose existence condition has been systematically understood. 
Towards a systematic understanding of the existence of the Cho--Maison monopole, in this paper, we explicitly construct a generalized Cho--Maison monopole configuration in a concrete setup. 
In particular, as a simple yet nontrivial example, we consider a model with gauge symmetry
$\mathrm{SU(3)_A} \times \mathrm{SO(3)_B} \to \mathrm{SO(3)}_{\mathrm{diag}}$,
which realizes a non-Abelian generalization of the diagonal breaking pattern.
This construction provides a concrete example suggesting that generalized Cho--Maison monopoles can arise in a broader class of theories with non-Abelian diagonal symmetry breaking, and offers useful insight into the conditions for the existence of generalized Cho--Maison monopoles \cite{SatoShimamoriMiya}. 

Second, the Cho--Maison monopole still contains an Abelian Dirac monopole component, and as a consequence, the energy density diverges at the origin. 
Various approaches have been proposed to render the Cho--Maison monopole energy finite (see, for example, refs.~\cite{Cho:2013vba, Ellis:2016glu, Arunasalam:2017eyu}). 
These works focus on the fact that the energy divergence originates from the $\mathrm{U(1)_{Y}}$ gauge field, and attempt to resolve it either by promoting the corresponding gauge coupling to a field-dependent effective coupling or by incorporating higher-dimensional operators. 
In this paper, we propose an alternative scenario in which the energy of the electroweak monopole is rendered finite by embedding it into an ’t~Hooft--Polyakov monopole in a UV theory. 
In such a framework, the Standard Model appears as a low energy effective theory of the UV theory, and the Cho--Maison monopole can be understood as a low energy effective description of a 't Hooft--Polyakov monopole. The stability guaranteed for the ’t~Hooft--Polyakov monopole in the UV theory is consequently inherited by the Cho--Maison monopole in the low-energy effective description. 

The organization of this paper is as follows. 
In section \ref{sec:electroweak monopole}, we briefly review the electroweak monopole configuration found by Cho and Maison \cite{Cho:1996qd}.
In section \ref{sec:generalization}, we present a concrete realization of a generalized Cho--Maison monopole in a toy model with symmetry breaking $\mathrm{SU(3)} \times \mathrm{SO(3)} \to \mathrm{SO(3)}_{\mathrm{diag}}$,
where we explicitly construct the monopole solution and show that the configuration can be smoothly described by gluing two gauge patches.
In section~\ref{sec:UV completion}, we then turn to the question of ultraviolet completion, and discuss how the electroweak monopole can be embedded into a framework in which a regular ’t~Hooft--Polyakov monopole arises, thereby providing a consistent resolution of the short-distance behavior.
Finally, we summarize our findings and discuss potential implications and future directions in section \ref{sec:summary}.

\section{Review of Electroweak Cho--Maison Monopole}\label{sec:electroweak monopole}
In this section, we briefly review the construction of a monopole configuration in the Standard Model found by Cho and Maison \cite{Cho:1996qd} (see also ref.~\cite{Ellis:2016glu}.)
The Lagrangian of the electroweak sector in the Standard Model is given by
\begin{align}
  \mathcal{L}
  &= \left| D_\mu H \right|^2
   - \frac{\lambda}{2}\left(H^\dagger H - \frac{\mu^2}{\lambda}\right)^2
   - \frac{1}{4} F^a_{\mu\nu} F^{a\,\mu\nu}
   - \frac{1}{4} B_{\mu\nu} B^{\mu\nu}, \label{CMLag}
\end{align}
where $H$ is the Higgs doublet field, and its covariant derivative is defined as
\begin{align}
   D_{\mu}H=\left(\partial_\mu - \mathrm{i}\, g \frac{\sigma^a}{2}  {A}^a_\mu - \mathrm{i}\, \frac{g'}{2} B_\mu\right)H.
\end{align}
Here $\sigma^a \,(a=1,2,3)$ are the Pauli matrices, and $A^a_\mu$ $(a=1,2,3)$ and $B_\mu$ denote the $\mathrm{SU(2)_{L}}$ and $\mathrm{U(1)_{Y}}$ gauge fields, respectively.
Let us decompose $H$ by using a real field $\rho$ and a complex unit vector field $\xi$ as
\begin{align}
  H(x) = \dfrac{v}{\sqrt{2}}\rho(x) \,\xi(x).
\end{align}
Substituting this into the Lagrangian \eqref{CMLag}, we obtain
\begin{align}
  \mathcal{L}
  = \frac{v^2}{2}(\partial_\mu \rho)^2 
     + \frac{v^2\rho^2}{2}\,|D_\mu \xi|^2 
     - \frac{\mu^2}{4}\left(\rho^2 - 1\right)^2  
     - \frac{1}{4}F_{\mu\nu}F^{\mu\nu} 
     - \frac{1}{4}B_{\mu\nu}B^{\mu\nu},
\end{align}
where $v = \sqrt{2}\mu / \sqrt{\lambda}$ is the VEV of the field $\rho$.

Let us choose the following ansatz for the fields in spherical coordinates $(t,r,\theta,\varphi)$:
\begin{equation}\label{CMansatz01}
\begin{aligned}
  \rho &= \rho(r), \quad
  \xi =  \begin{pmatrix} -e^{-\mathrm{i}\varphi}\sin(\theta/2) \\ \cos(\theta/2) \end{pmatrix}, \\
  A_0^a &= - \frac{1}{g} A(r) \hat r_a, \quad
  A_i^a = \frac{1}{g} (f(r) - 1) f_{abc} \hat r_b \partial_i \hat r_c, \\
  B_0 &= -\frac{1}{g'} B(r), \quad
  B_i = -\frac{1}{g'} (1-\cos\theta) \partial_i \varphi,
\end{aligned}
\end{equation}
where $\hat r = (\sin\theta \cos\varphi,\, \sin\theta \sin\varphi,\, \cos\theta)$ is the unit radial vector in $\mathbb{R}^3$.
Note that $\xi(\theta,\varphi)$ can be expressed as
\begin{align}
    \xi(\theta,\varphi) = U_2(\theta,\varphi) \left(\begin{array}{c} 0 \\ 1 \end{array}\right),
\end{align}
where $U_2(\theta,\varphi)$ is defined as
\begin{align}
    U_2(\theta,\varphi) 
    &= \exp\left( -\mathrm{i} \varphi \sigma^3/2 \right)
       \exp\left( -\mathrm{i} \theta \sigma^2/2 \right)
       \exp\left( \mathrm{i} \varphi \sigma^3/2 \right). \label{eq:u2}
\end{align}
The field configuration given in eq.~\eqref{CMansatz01} is ill-defined at $\theta = \pi$. This implies that the field on a sphere surrounding the Cho--Maison monopole cannot be described within a single gauge patch.

A well-defined description of this monopole can be given by using two gauge patches.
By using a $\mathrm{U(1)_Y}$ gauge transformation, 
\begin{align}
    H \to \exp\left( \frac{\i}{2}\alpha(x) \right) H, \quad
    B_\mu \to B_\mu + \frac{1}{g'} \partial_\mu \alpha(x),
\end{align}
with $\alpha(x) = 2\varphi$, 
the field configuration given in eq.~\eqref{CMansatz01} is transformed to
\begin{equation}\label{CMansatz02}
\begin{aligned}
  \rho &= \rho(r), \quad
  \xi =
  \begin{pmatrix}
    -\sin(\theta/2) \\
    e^{\mathrm{i}\varphi} \cos(\theta/2)
  \end{pmatrix}, \\
  A_0^a &= - \frac{1}{g} A(r) \hat r_a, \quad
  A_i^a = \frac{1}{g} (f(r) - 1)
  f_{abc} \hat r_b \partial_i \hat r_c, \\
  B_0 &= -\frac{1}{g'} B(r), \quad
  B_i = \frac{1}{g'} (1+\cos\theta)\partial_i\varphi .
\end{aligned}
\end{equation}
In this gauge, the configuration at $\theta = \pi$ is well-defined while ill-defined at $\theta = 0$.
Thus, we can use eq.~\eqref{CMansatz01} for the description in the northern hemisphere ($0 \leq \theta \leq \pi/2)$ and eq.~\eqref{CMansatz02} for the southern hemisphere ($\pi/2 \leq \theta \leq \pi$).
This property is analogous to the Wu--Yang monopole construction \cite{Wu:1976ge}, rather than the 't~Hooft--Polyakov monopole \cite{tHooft:1974kcl, Polyakov:1974ek}.

As discussed in ref.~\cite{Cho:1996qd},
once we identify field configurations related by $\mathrm{U(1)_Y}$ gauge transformations, the target space of $\xi$ is given by $\mathbb{C}\mathrm{P}^{1} \simeq S^2$.
Thus, the angular dependence of $\xi$ in eq.~\eqref{CMansatz01} can be interpreted as a nontrivial element of $\pi_2(\mathbb{C}\mathrm{P}^{1}) = \mathbb{Z}$.
It is worth emphasizing that the gauge and scalar field ansatz are formally identical to those used in the construction of the Nambu monopole \cite{Nambu:1977ag}. In the Nambu monopole, the Dirac string is promoted to a physical $Z$-flux tube carrying energy, and the monopole appears as its endpoint. In contrast, in the Cho--Maison monopole, the string-like singularity is removed by patching gauge charts, and thus remains a gauge artifact rather than a physical object.

From the ansatz given in eq.~\eqref{CMansatz01}, we obtain a monopole/dyon solution in the electroweak theory, dressed by the $W$ and $Z$ bosons and the Higgs field, which reduces to a monopole for $A=B=0$.
The presence of the time components in the ansatz gives rise to an electrically charged (dyonic) solution.
With this ansatz, the equations of motion become
\begin{subequations} \label{eq:eom CM monopole}
  \begin{align}
  \dv[2]{\rho}{r}
  + \frac{2}{r} \dv{\rho}{r}
  - \frac{f^2}{2r^2}\,\rho
  + \frac{(A-B)^2}{4}\,\rho
  - \mu^2\!\left(\rho^2-1\right)\!\rho
  &= 0 , \\
  \dv[2]{f}{r}
  - \frac{f}{r^2}(f^2-1)
  - \left(\frac{g^2v^2}{4}\rho^2 - A^2\right) f
  &= 0 , \\
  \dv[2]{A}{r}
  + \frac{2}{r} \dv{A}{r}
  - \frac{2f^2}{r^2}\,A
  - \frac{g^2v^2}{4}\rho^2\,(A-B)
  &= 0 , \\
  \dv[2]{B}{r}
  + \frac{2}{r} \dv{B}{r}
  + \frac{g'^2v^2}{4}\rho^2\,(A-B)
  &= 0.
  \end{align}
\end{subequations}
After an appropriate unitary gauge transformation $U$ such that $\xi \;\rightarrow\; U\xi = (0~1)^T$, one may obtain
\begin{align}
    W_\mu^1 &= \frac{f(r)}{g} \left(-\sin\varphi \partial_\mu\theta - \sin\theta\cos\varphi \partial_\mu \varphi \right), \\
    W_\mu^2 &= \frac{f(r)}{g} \left(\cos\varphi \partial_\mu \theta -\sin\theta\sin\varphi \partial_\mu\varphi \right), \\
    W_\mu^3 &= -\frac{1}{g} A(r) \delta_{\mu 0}  -\frac{f(r)}{g} \left( 1 - \cos\theta\right) \partial_\mu\varphi.
\end{align}
and obtain the physical gauge fields by rotating through the electroweak mixing angle $\theta_W$:
\begin{equation}\label{ans2}
\begin{aligned}
  W_{\mu}
  &=\dfrac{\mathrm{i}}{g}
  \frac{f(r)}{\sqrt2}
  e^{\mathrm{i}\varphi}
  \left(
    \partial_\mu \theta
    + \mathrm{i}\sin\theta\, \partial_\mu \varphi
  \right), \\
  A_{\mu}^{\rm EM}
  &=- e
  \left(
    \frac{1}{g^2}A(r)
    + \frac{1}{g'^2} B(r)
  \right)
  \partial_{\mu}t
  -\frac{1}{e}(1-\cos\theta)\partial_{\mu}\varphi, \\
  Z_{\mu}
  &=- \frac{e}{gg'}
  \big(A(r)-B(r)\big)
  \partial_{\mu}t .
\end{aligned}
\end{equation}
where
\begin{align}
    W_\mu \equiv \frac{1}{\sqrt{2}}( A_\mu^1 + \mathrm{i} A_\mu^2 ) , \quad
    \left(\begin{array}{cc}
        A_\mu^\mathrm{EM} \\
        Z_\mu
    \end{array}\right)
    \equiv
    \left(\begin{array}{cc}
        \cos\theta_W & \sin\theta_W \\
        -\sin\theta_W & \cos\theta_W
    \end{array}\right)
    \left(\begin{array}{cc}
        B_\mu \\
        A_\mu^3
    \end{array}\right),
\end{align}
and the electric charge is given by $e = g \sin \theta_W = g^\prime \cos \theta_W$.
The simplest nontrivial solution to the equations of motion \eqref{eq:eom CM monopole} is
\begin{align}
    A(r) = B(r) = f(r) = 0, \quad
    \rho(r) = 1,
\end{align}
which describes a point monopole with magnetic charge $4\pi/e$:
\begin{equation*}
  A_\mu^\text{EM} = -\frac{1}{e}(1 - \cos \theta)\partial_\mu\varphi \, .
\end{equation*}
In this solution, both $\mathrm{SU(2)_L}$ and $\mathrm{U(1)_Y}$ gauge field exhibit singular behavior at $r=0$.
Now let us require a regular behavior of $\mathrm{SU(2)_L}$ gauge field as Cho and Maison \cite{Cho:1996qd}. Then, one possible set of boundary conditions is
\begin{eqnarray}
  &\rho(0)=0,~~f(0)=1,~~A(0)=0,~~B(0)=b_0, \nonumber\\
  &\rho(\infty)=v ,~f(\infty)=0,~A(\infty)=A_0,~B(\infty)=B_0,\label{bc0}
\end{eqnarray}
where $0 \leq A_0 \leq e v$ and $0 \leq b_0 \leq A_0$.
This boundary condition gives more general electroweak monopole/dyon solutions.
As discussed in ref.~\cite{Cho:1996qd}, the electric and magnetic charge of the Cho--Maison dyon is given by
\begin{align}
  &q_e=-\dfrac{8\pi}{e}\sin^2\theta_W \int_0^\infty f^2 A \,\dd{r} 
  =\dfrac{4\pi}{e} A_1, \\
  &q_m = \dfrac{4\pi}{e},\label{eq:Charge}
\end{align}
where \(A_1\) is the coefficient of the \(1/r\) term in the asymptotic expansion
\[
  A(r) \xrightarrow{r\to\infty} A_0 + \frac{A_1}{r} + \cdots .
\]
Here \(A_0\) specifies the asymptotic value of the electric potential and is used as a boundary condition for the dyon solution, while \(A_1\) determines the Coulombic tail of the electric field and hence the electric charge.

In contrast to the ’t~Hooft--Polyakov monopole, the Cho--Maison monopole exhibits a singular behavior at the origin.
Around the origin $r\simeq 0$, the equation of motion for $\rho$ given in eq.~\eqref{eq:eom CM monopole} behaves as
\begin{align}
  \dv[2]{\rho}{r}
  + \frac{2}{r} \dv{\rho}{r}
  - \frac{\rho}{2r^2}
  \simeq 0.
\end{align}
Here we have used $f\simeq 1$ as indicated in eq.~\eqref{bc0}. The solution to this equation satisfying the boundary condition in eq.~\eqref{bc0} behaves as
\begin{align}
    \rho \propto r^{\tfrac{\sqrt{3}-1}{2}}. \label{eq:CMmonopole at the origin}
\end{align}
Thus, $\rho$ is not an analytic function of $r$ around the origin.
Moreover, the Cho--Maison electroweak monopole and dyon~\cite{Cho:1996qd} suffer from 
a divergence in the energy density at  $r \to 0$.
Let us evaluate the total energy $E$ of the dyon configuration, which has the form~\cite{Cho:2013vba}:
\begin{align}
  E &= E_0 +E_1,  \nonumber \\
  E_0 &= 4\pi\int_0^\infty \frac{\dd{r}}{2 r^2}
  \bigg\{\frac{1}{g'^2}+ \frac1{g^2}(f^2-1)^2\bigg\}, \nonumber\\
  E_1 &= 4\pi \int_0^\infty \dd{r} \bigg\{\frac{v^2}{2} \left(r \dv{\rho}{r}\right)^2
  +\frac1{g^2} \left( \left(\dv{f}{r}\right)^2 +\frac{1}{2}\left(r \dv{A}{r} \right)^2 + f^2 A^2 \right) \nonumber\\
  &\quad\quad\quad +\frac{1}{2g'^2} \left(r \dv{B}{r} \right)^2 
  +\frac{\mu^2 r^2}{4}\big(\rho^2-1 \big)^2 +\frac{v^2}{4} f^2\rho^2
  +\frac{v^2r^2}{8} (B-A)^2 \rho^2 \bigg\}.
\end{align}
We see that, with the boundary conditions given in eq.~\eqref{bc0}, $E_1$ is finite, whereas the first term in $E_0$ diverges at the origin. 
Thus, the energy diverges due to the $\mathrm{U(1)_Y}$ sector, motivating a mechanism for finite-energy regularization.

\begin{figure}[h]
  \centering
  \includegraphics[width=0.7\textwidth]{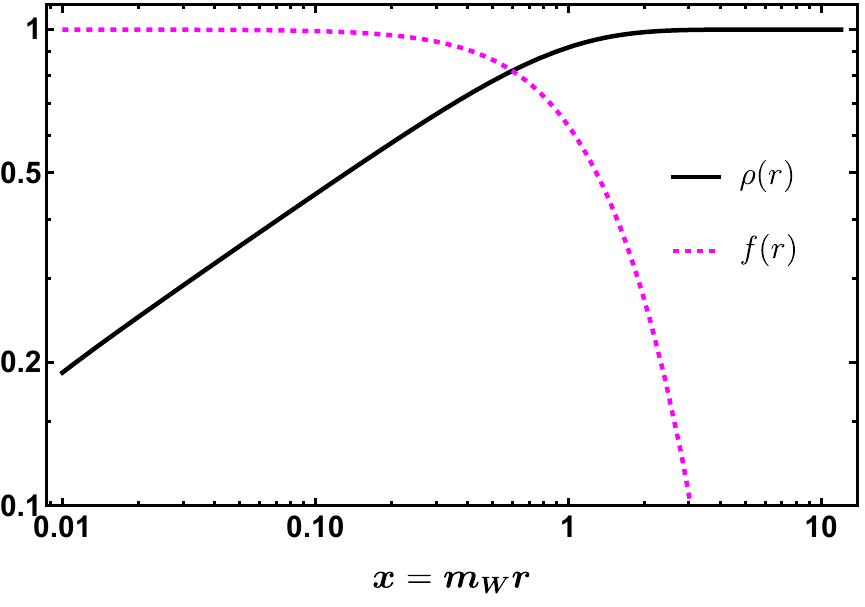}
  \caption{The profile functions of the Cho--Maison monopole in the electroweak $\mathrm{SU(2)_L} \times \mathrm{U(1)_Y}$ theory. The black solid and magenta dashed curves correspond to $\rho(r)$ and $f(r)$, respectively. 
  We use the Standard Model gauge couplings as $g=0.65$ and $g^\prime=0.35$. For the Higgs potential, we take $\lambda = m_H^2/v^2$ and $\mu = m_H/\sqrt{2}$ where $v=246\,\mathrm{GeV}$ and $m_H=125\,\mathrm{GeV}$. 
  We express the equations in dimensionless form by rescaling the radial coordinate as $x = m_W r$ with $m_W = gv/2$, and normalizing the scalar fields by the vacuum expectation value $v$.}
  \label{fig:CMdyon}
\end{figure}

Figure \ref{fig:CMdyon} shows the profile functions obtained by solving eq.~\eqref{eq:eom CM monopole} with the boundary conditions given in eq.~\eqref{bc0}.
By setting $A(r) = B(r) = 0$, we restrict ourselves to a purely magnetic sector, and hence the configuration represents a monopole solution.
The scalar field $\rho(r)$ approaches its vacuum expectation value at large distances, while the gauge field profile $f(r)$ rapidly decays to zero. 
A notable difference from the 't Hooft--Polyakov monopole appears in the power-law behavior near the origin. 
While the scalar field in the 't Hooft--Polyakov solution behaves as $\rho(r) \propto r$, 
in the present case, it follows a non-integer power law given in eq.~\eqref{eq:CMmonopole at the origin}. 

\section{Generalized Cho--Maison monopole}\label{sec:generalization}
In the previous section, we discussed the Cho--Maison monopole configuration~\cite{Cho:1996qd} in the electroweak theory, where the gauge symmetry $\mathrm{SU(2)_L} \times \mathrm{U(1)_Y}$ is broken to $\mathrm{U(1)_{EM}}$ by the Higgs doublet field. 
In this section, we discuss generalizations of this monopole configuration.
The essential structure of the Cho--Maison monopole is based on the electroweak symmetry breaking pattern
\begin{align}
    \mathrm{SU(2)} \times \mathrm{U(1)}
    \;\longrightarrow\;
    \mathrm{U(1)}.
\end{align}
Therefore, similar monopole configurations are expected to exist in more general gauge theories containing an $\mathrm{SU(2)} \times \mathrm{U(1)}$ subgroup with the same symmetry breaking structure.

We now consider a gauge theory with
\begin{align}
    G=G_{\rm A}\times G_{\rm B},
\end{align}
which is spontaneously broken to an unbroken subgroup $H$ by a Higgs VEV
$\Phi_0$.  We assume that $G_{\rm A}$ is simply connected and that
$G_{\rm B}$ is a connected compact Lie group with nontrivial fundamental group.
Thus,
\begin{align}
    \pi_1(G_{\rm A})=0,\quad\pi_1(G_{\rm B})\neq0.
\end{align}
We focus on symmetry-breaking patterns for which the full vacuum manifold does
not support an ordinary 't Hooft--Polyakov monopole.  Our goal is to isolate
the remaining topology responsible for Cho--Maison-type configurations in this
setting.  The full vacuum manifold is
\begin{align}
    \mathcal M_{\rm vac}
    =\frac{G_{\rm A}\times G_{\rm B}}{H}.
\end{align}
Since $\pi_2(G_{\rm A}\times G_{\rm B})=0$, the long exact sequence gives
\begin{align}
    \pi_2(\mathcal M_{\rm vac})
    \simeq
    \ker\!\left[\pi_1(H)
    \to\pi_1(G_{\rm A}\times G_{\rm B})\right].
\end{align}
The absence of such ordinary topological monopoles means that the above kernel
is trivial, and hence
\begin{align}
    \pi_2(\mathcal M_{\rm vac})=0.
\end{align}
Thus the Cho--Maison-type monopole is not an ordinary topological monopole
classified by the full vacuum manifold.  We then identify the remaining
topological data with the
winding of the Higgs direction after quotienting out a redundancy associated
with the $G_{\rm B}$ gauge symmetry.

To make this statement precise, let us define the subgroup of $G_{\rm A}$
whose action on the VEV can be compensated by a $G_{\rm B}$ gauge
transformation,
\begin{align}
  K_{\rm A}
  =\left\{g_{\rm A}\in G_{\rm A}\,\middle|\,\exists g_{\rm B}\in G_{\rm B}
  \ \text{such that}\,(g_{\rm A},g_{\rm B})\Phi_0=\Phi_0\right\}.
\end{align}
Elements of $K_{\rm A}$ act trivially on the Higgs direction once
$G_{\rm B}$ gauge transformations are allowed.
Therefore, when we classify Higgs directions generated by the action of
$G_{\rm A}$ on $\Phi_0$, two elements of $G_{\rm A}$ should be identified if
they differ by an element of $K_{\rm A}$.  Equivalently, they are identified if
they induce the same Higgs direction after allowing a compensating
$G_{\rm B}$ gauge transformation.  The resulting space of Higgs directions is
\begin{align}
  \mathcal M_{\rm H}
  \equiv\frac{G_{\rm A}}{K_{\rm A}}.\label{eq:DefofCMVaccum}
\end{align}
Since $G_{\rm A}$ is simply connected, the long exact sequence associated with $K_{\rm A}\to G_{\rm A}\to G_{\rm A}/K_{\rm A}$ gives
\begin{align}
\pi_2(\mathcal M_{\rm H})
=
\pi_2\left(\frac{G_{\rm A}}{K_{\rm A}}\right)
\simeq
\pi_1(K_{\rm A}).
\end{align}
Therefore the essential condition for the existence of Cho--Maison-type monopole configurations is not that $K_{\rm A}$ contains a $\mathrm{U}(1)$ factor, but rather that
\begin{align}
  \pi_1(K_{\rm A})\neq 0.
\end{align}

In the electroweak theory, this construction reduces to the familiar
description in terms of the direction of the Higgs doublet, where doublets
differing only by an overall phase are identified.  Taking
\(\xi_0=(0~1)^T\), an \(\mathrm{SU(2)_L}\) transformation maps
\(\xi_0\) to \(\xi\), while the subgroup
\(K_{\rm A}=\mathrm{U(1)}_{T^3}\subset \mathrm{SU(2)_L}\) acts only by an
overall phase on \(\xi_0\), which can be compensated by a
\(\mathrm{U(1)_Y}\) gauge transformation.  Thus the relevant variable is not
\(\xi\) itself but its equivalence class
\begin{align}
  [\xi]=\left\{e^{\mathrm{i}\alpha}\xi\,\middle|\,\alpha\in \mathbb R/2\pi\mathbb Z\right\},
\end{align}
and the space of such classes is
\begin{align}
  \mathcal{M}_H = \mathrm{SU(2)_L}/\mathrm{U(1)}_{T^3}\simeq \mathbb{C}\mathrm{P}^1.
\end{align}
In this way, the space \(\mathcal M_{\rm H}\) defined in
eq.~\eqref{eq:DefofCMVaccum} provides the appropriate target space for
discussing the existence condition of Cho--Maison-type monopole configurations.

One of the most natural generalizations of the Cho--Maison monopole arises in $\mathrm{SU}(N)_{\mathrm{A}} \times \mathrm{U(1)_B}$ gauge theory with a Higgs field $H_N$ in the fundamental representation of $\mathrm{SU}(N)_{\mathrm{A}}$ carrying a nonzero $\mathrm{U(1)_B}$ charge. 
In this setup, $\mathrm{SU(2)_A} \subset \mathrm{SU}(N)_{\mathrm{A}}$ and $\mathrm{U(1)_B}$ play roles analogous to those of $\mathrm{SU(2)_L}$ and $\mathrm{U(1)_Y}$ in the electroweak theory. 
The VEV of $H_N$ then breaks
\begin{align}
    \mathrm{SU}(N)_{\mathrm{A}} \times \mathrm{U(1)_B}
    \;\longrightarrow\;
    \mathrm{SU}(N-1)_{\mathrm{A}} \times \mathrm{U(1)_{diag}},
\end{align}
where $\mathrm{U(1)_{diag}}$ is the diagonal subgroup of $\mathrm{U(1)_A}\subset \mathrm{SU}(N)_{\mathrm{A}}$ and $\mathrm{U(1)_B}$.
Then, the ansatz in eq.~\eqref{CMansatz01} can be naturally generalized for $H_N$ as
\begin{align}
    H_N(x) = \rho(r) \xi_N(\theta,\varphi), \quad
    \xi_N(\theta,\varphi) = \begin{pmatrix} -e^{-\mathrm{i}\varphi}\sin(\theta/2) \\ \cos(\theta/2) \\
  0 \\
  \vdots \\
  0 \end{pmatrix}.
\end{align}
Similar to the Cho--Maison electroweak monopole, once we identify field configuration related by $\mathrm{U(1)_B}$ gauge transformation, the target space of $\xi_N$ becomes $\mathbb{C}\mathrm{P}^{N-1}$ and the angular dependence of $\xi_N$ can be understood as a nontrivial element of $\pi_2(\mathbb{C}\mathrm{P}^{N-1}) = \mathbb{Z}$.
The ansatz for $\mathrm{SU(N)_A}$ and $\mathrm{U(1)_B}$ gauge fields is also a natural embedding of the Cho--Maison monopole ansatz. 
In the remainder of this section, we will discuss another yet nontrivial generalization of the Cho--Maison monopole.

Now we focus on a setup with $\mathrm{SU(3)_A} \times \mathrm{SO(3)_B} $ gauge symmetry where $\mathrm{SU(3)_A}$ and $\mathrm{SO(3)_B}$ are broken diagonally, and the symmetry breaking pattern is given by $\mathrm{SU(3)_A} \times \mathrm{SO(3)_B} \to \mathrm{SO(3)_{diag}}$.
The vacuum manifold is given by
\begin{align}
  \mathcal{M} = \frac{\mathrm{SU(3)_A} \times \mathrm{SO(3)_B}}{\mathrm{SO(3)_{diag}}} \simeq \mathrm{SU(3)} 
\end{align}
which has a trivial second homotopy group $\pi_2(\mathrm{SU(3)}) = 0$. Thus, there is no 't Hooft--Polyakov monopole solution. However, we show that a Cho--Maison-like monopole solution can be constructed, and the energy of this monopole solution also diverges at the origin.

As a concrete setup, we introduce
a Higgs field $\Phi$ transforming in the $(3,3)$ representation.
Under a gauge transformation, $\Phi$ transforms as
\begin{align}
    \Phi \to U \Phi R^T
\end{align}
with $U \in \mathrm{SU(3)_A}$ and $R \in \mathrm{SO(3)_B}$.
The Lagrangian of the model is given by
\begin{align}
  \mathcal{L}
  &=-\frac{1}{4}\, G^A_{\mu\nu} G^{A\,\mu\nu}
  -\frac{1}{4}\, W^a_{\mu\nu} W^{a\,\mu\nu}
  +\mathrm{Tr}\!\left[(D_\mu \Phi)^\dagger (D^\mu \Phi)\right]
  - V(\Phi),\\
  D_\mu \Phi
  &=\partial_\mu \Phi
  - \mathrm{i} g\, G_\mu^A \frac{\lambda^A}{2}\, \Phi
  + \mathrm{i} g'\, \Phi\, W_\mu^a J^a,\\
  G^A_{\mu\nu}
  &=\partial_\mu G^A_\nu - \partial_\nu G^A_\mu
  + g f^{ABC} G^B_\mu G^C_\nu,\\
  W^a_{\mu\nu}
  &=
  \partial_\mu W^a_\nu - \partial_\nu W^a_\mu
  + g^\prime \varepsilon^{abc} W^b_\mu W^c_\nu,
\end{align}
where $\lambda^A~(A=1,\ldots,8)$ denote the Gell--Mann matrices, and $J^a~(a=1,2,3)$ are the generators of $\mathrm{SO(3)_B}$ in the fundamental (vector) representation, which is defined as
\begin{align}
    J^1 &= \begin{pmatrix}
        0 & 0 & 0 \\
        0 & 0 & -\mathrm{i} \\
        0 & \mathrm{i} & 0
    \end{pmatrix}, \quad
    J^2 = \begin{pmatrix}
        0 & 0 & \mathrm{i} \\
        0 & 0 & 0 \\
        -\mathrm{i} & 0 & 0
    \end{pmatrix}, \quad
    J^3 = \begin{pmatrix}
        0 & -\mathrm{i} & 0 \\
        \mathrm{i} & 0 & 0 \\
        0 & 0 & 0
    \end{pmatrix}.
\end{align}

\subsection{Potential for \texorpdfstring{\(\Phi\)}{Phi} and Its VEV}
We take $V(\Phi)$ as the most general renormalizable scalar potential consistent with the $\mathrm{SU(3)_A} \times \mathrm{SO(3)_B}$ symmetry as
\begin{align}
  V(\Phi)
   &=
   \frac{\lambda}{2}\bigl({\rm tr}[\Phi \Phi^\dagger]\bigr)^2
   + \frac{\lambda'}{2}\,{\rm tr}[\Phi \Phi^\dagger \Phi \Phi^\dagger]
   - \mu^2\, {\rm tr}[\Phi \Phi^\dagger]
   - \kappa \bigl( \det\Phi + \det\Phi^\dagger \bigr). \label{eq:Vphi}
\end{align}
In the following analysis, we take $\lambda > 0$, $\lambda' > 0$, $\mu^2>0$, and $\kappa \ge0$.
In general, 
by the singular value decomposition, any complex \(3\times 3\) matrix \(\Phi\) can be written as
\begin{align}
   \Phi = U_{\mathrm{L}} \,\Sigma\, U_{\mathrm{R}}^\dagger,
\end{align}
where $U_{\mathrm{L}}$ and $U_{\mathrm{R}}$ are $3\times 3$ unitary matrices, and
\begin{align}
   \Sigma = {\rm diag}(\sigma_1,\sigma_2,\sigma_3),
   \quad
   \sigma_i \ge 0.
\end{align}
The first three terms in eq.~\eqref{eq:Vphi} do not depend on $U_{\mathrm{L}}$ and $U_{\mathrm{R}}$, and the last term is proportional to $\det U_{\mathrm{L}} U_{\mathrm{R}}^\dagger + h.c.$.
Thus, for given $\Sigma$, $V(\phi)$ is minimized if $\det U_{\mathrm{L}} U_{\mathrm{R}}^\dagger = 1$.
 Hence, one may, without loss of generality, choose a basis in which the VEV is diagonal:
\begin{align}
   \ev{\Phi} = {\rm diag}(v_1,v_2,v_3),
\end{align}
with \(v_i\) taken real. 
For $\lambda' > 0$ and $\mu^2 > 0$, $v_1$, $v_2$, and $v_3$ are equal at the global minimum of $V(\Phi)$. The VEV of $\Phi$ is given as
\begin{align}
   \ev{\Phi} = v\, I_3, \label{eq:VEV phi}
\end{align}
where
\begin{align}
   v
   =\frac{\kappa + \sqrt{\kappa^2 + 4(3\lambda+\lambda')\mu^2}}{2(3\lambda+\lambda')}.
\end{align}
Note that the scalar potential $V(\Phi)$ given in eq.~\eqref{eq:Vphi} is the same as the potential of $\mathrm{SU(3)} \times \mathrm{SU(3)}$ linear sigma model and the symmetry breaking pattern in this potential has been studied in ref.~\cite{Bai:2017zhj}.
The VEV given in eq.~\eqref{eq:VEV phi} preserves the diagonal subgroup of \(\mathrm{SU(3)_A}\times \mathrm{SO(3)_B}\), and this vacuum configuration determines the asymptotic boundary condition for the scalar field in the monopole ansatz.

\subsection{Monopole Ansatz}
Let us discuss an ansatz for monopole configuration \`a la Cho--Maison.
The ansatz can be obtained by assuming that $\mathrm{SU(2)_A} \subset \mathrm{SU(3)_A}$ and $\mathrm{U(1)_B} \subset \mathrm{SO(3)_B}$ have similar roles as $\mathrm{SU(2)_L}$ and $\mathrm{U(1)_Y}$ in the Cho--Maison monopole in the electroweak theory \cite{Cho:1996qd}.
For the $\mathrm{SU(3)_A}$ gauge field, we take
\begin{align}
    G_0^a = -\frac{1}{g} A(r) \hat r^a, 
    \quad G_i^a = \frac{1}{g} \left[ f(r) - 1 \right]
    f^{abc} \hat r^b \partial_i \hat r^c
    ,\quad a=1,2,3
    \label{eq:su3 gauge field ansatz}
\end{align}
while all other components vanish:
\begin{align}
    G_\mu^a =0,\quad a=4,\dots,8.
\end{align}
For the $\mathrm{SO(3)_B}$ gauge field, we assume
\begin{align}
    W_0^3 = -\frac{1}{2g'} B(r), \quad
    W_i^3 = -\frac{1}{2g'}
    \left(1-\cos\theta\right)\partial_i\varphi,\label{eq:so3 gauge field ansatz}
\end{align}
while the remaining components vanish:
\begin{align}
    W_\mu^1=W_\mu^2=0.
\end{align}
In the positive $z$ direction, the monopole has a magnetic field in a direction of $g\, G_\mu^3 + g'\, W_\mu^3$. Thus, the corresponding gauge symmetry is unbroken and $\langle\Phi\rangle$ in this direction should satisfy
\begin{align}
    T^3 \langle \Phi \rangle + \langle \Phi \rangle J^3 = 0.
\end{align}
Since $\langle \Phi \rangle$ can be transformed to the VEV in eq.~\eqref{eq:VEV phi} by a $\mathrm{SU(3)_A} \times \mathrm{SO(3)_B}$ gauge transformation, $\langle \Phi \rangle / v$ should be a unitary matrix in the limit of $r\to\infty$. Since $U(\theta,\varphi)$ defined in eq.~\eqref{eq:u3} is $I_3$ in the positive $z$ direction ($\theta = 0$), we take
\begin{align}
  \langle \Phi \rangle
  = v
  \begin{pmatrix}
        \frac{1}{\sqrt{2}} & -\frac{\mathrm{i}}{\sqrt{2}} & 0 \\
        -\frac{\mathrm{i}}{\sqrt{2}} & \frac{1}{\sqrt{2}} & 0 \\
        0 & 0 & 1
    \end{pmatrix}
    \label{SU3SO3VEVPhi}
\end{align}
for the positive $z$ direction with $r\to\infty$.
Thus, we adopt the following ansatz for the scalar field:
\begin{align}
    \Phi(x) = v\,U_3(\theta,\varphi)\,\bigl[\rho(r)\, T_\rho + \chi(r)\, T_\chi \bigr], \label{eq:phi ansatz}
\end{align}
where $U_3(\theta,\varphi)$, $T_\rho$, and $T_\chi$ are defined as
\begin{align}
    U_3(\theta,\varphi) 
    &= \exp\left( -\mathrm{i} \varphi \lambda^3/2 \right)
       \exp\left( -\mathrm{i} \theta \lambda^2/2 \right)
       \exp\left( \mathrm{i} \varphi \lambda^3/2 \right), \label{eq:u3}\\
    T_\rho
    &=\frac{1}{\sqrt{2}}
    \begin{pmatrix}
        1 & -\mathrm{i} & 0 \\
        -\mathrm{i} & 1 & 0 \\
        0 & 0 & 0
    \end{pmatrix},
    \quad
    T_\chi
    =
    \begin{pmatrix}
        0 & 0 & 0\\
        0 & 0 & 0\\
        0 & 0 & 1
    \end{pmatrix}. \label{eq:Trho Tchi}
\end{align}
$U_3(\theta,\varphi)$ expresses an angular-dependent $\mathrm{SU(2)_A}~(\subset \mathrm{SU(3)_A})$ transformation to realize a spherical configuration as $U_2(\theta,\varphi)$ in eq.~\eqref{eq:u2}.
The behavior of $\rho(r)$ and $\chi(r)$ at $r\to\infty$ are given as
\begin{align}
    \rho(r) \to 1,\quad \chi(r) \to 1\quad(r\to\infty),
\end{align}
By using eqs.~\eqref{eq:u3} and \eqref{eq:Trho Tchi},
the ansatz in eq.~\eqref{eq:phi ansatz} can be explicitly written as
\begin{align}
   \frac{\langle \Phi \rangle}{v}  =
    \frac{\rho(r)\cos(\theta/2)}{\sqrt{2}}
    \begin{pmatrix}
        1 & \i & 0\\
        \i & 1 & 0\\
        0 & 0 & 0
    \end{pmatrix}
    + \frac{\rho(r)\sin(\theta/2)}{\sqrt{2}}
    \begin{pmatrix}
        \i e^{-\mathrm{i}\varphi} & - e^{-\mathrm{i}\varphi} & 0\\
        e^{\mathrm{i}\varphi} & -\i e^{\mathrm{i}\varphi} & 0\\
        0 & 0 & 0
    \end{pmatrix}
    + \chi(r) 
    \begin{pmatrix}
        0 & 0 & 0\\
        0 & 0 & 0\\
        0 & 0 & 1
    \end{pmatrix}. \label{eq:phi ansatz 2}
\end{align}

Similar to the electroweak Cho--Maison monopole given in eq.~\eqref{CMansatz01}, this field configuration is ill-defined at $\theta = \pi$. However, this ill-definedness at $\theta = \pi$ can be absorbed by a gauge transformation.
By using a $\mathrm{SO(3)_B}$ gauge transformation 
\begin{align}
\langle \Phi \rangle \to \langle \Phi \rangle \begin{pmatrix}
    \cos\alpha(x) & -\sin\alpha(x) & 0 \\
    \sin\alpha(x) & \cos\alpha(x) & 0 \\
    0 & 0 & 1
\end{pmatrix}, \quad
    W_\mu^3 \to W_\mu^3 + \frac{1}{g'} \partial_\mu\alpha(x),
\end{align}
with $\alpha(x) = \varphi$, $\langle \Phi \rangle$ and $W_i^3$ given in eqs.~\eqref{eq:phi ansatz 2} and \eqref{eq:so3 gauge field ansatz} is transformed to
\begin{align}
    \frac{\langle \Phi \rangle}{v} 
    &\to 
    \frac{\rho(r)\cos(\theta/2)}{\sqrt{2}}\begin{pmatrix}
        e^{\mathrm{i}\varphi} & \i e^{\mathrm{i}\varphi} & 0\\
        \i e^{-\mathrm{i}\varphi} & e^{-\mathrm{i}\varphi} & 0\\
        0 & 0 & 0
    \end{pmatrix}
    + \frac{\rho(r)\sin(\theta/2)}{\sqrt{2}}\begin{pmatrix}
        \i & - 1 & 0\\
        1 & -\i  & 0\\
        0 & 0 & 0
    \end{pmatrix}
    + \chi(r) \begin{pmatrix}
        0 & 0 & 0\\
        0 & 0 & 0\\
        0 & 0 & 1
    \end{pmatrix}, \\
    W_i^3 &\to \frac{1}{2g'}(1+\cos\theta)\partial_i\varphi. \label{eq:phi ansatz southernhemisphere}
\end{align}
In this gauge, the configuration at $\theta = \pi$ is well-defined while ill-defined at $\theta = 0$.
Thus, we could use eqs.~\eqref{eq:phi ansatz 2} and \eqref{eq:so3 gauge field ansatz} for the description in the northern hemisphere ($0 \leq \theta \leq \pi/2)$ and eq.~\eqref{eq:phi ansatz southernhemisphere} for the southern hemisphere ($\pi/2 \leq \theta \leq \pi$).

As indicated in eq.~\eqref{eq:so3 gauge field ansatz}, $\mathrm{SO(3)_B}$ gauge field has a singular behavior at the origin. This is similar to the electromagnetic gauge field in the Dirac/Wu--Yang monopole and $\mathrm{U(1)_Y}$ gauge field in the electroweak Cho--Maison monopole.
On the other hand, we assume $\mathrm{SU(3)_A}$ gauge field is regular at the origin as $\mathrm{SU(2)_L}$ gauge field in the electroweak Cho--Maison monopole.
To guarantee this behavior, we impose a boundary condition of $f(r)$ at the origin as
\begin{align}
    f(0) = 1.
\end{align}
To be consistent with this boundary condition and the equation of motion, we impose a boundary condition on $\rho(r)$ as
\begin{align}
    \rho(0) = 0.
\end{align}
while $\chi(r)$ at the origin can be a nonzero value as
\begin{align}
    \chi(0) = \chi_0.
\end{align}

Note that once we identify field configuration related by $\mathrm{SO(3)_B}$ gauge transformation, the VEV $\langle \Phi \rangle $ at $r=\infty$ parametrizes ${\rm SU(3)}/{\rm SO(3)}$ and the angular dependence of the ansatz can be understood as a nontrivial element of $\pi_2( \mathrm{SU(3)}/\mathrm{SO(3)} ) = \mathbb{Z}_2$.

\subsection{Equations of Motion and Numerical Results}
By using the ansatz given in eqs.~\eqref{eq:su3 gauge field ansatz}, \eqref{eq:so3 gauge field ansatz}, and  \eqref{eq:phi ansatz}, 
we obtain the following equation of motion for $\rho,\chi,A, B$, and $f$:
\begin{subequations}\label{SU3SO3EoM}
\begin{align}
  \dv[2]{\rho}{r}
  + \frac{2}{r}\,\dv{\rho}{r}
  - \frac{f^2}{2r^2}\,\rho
  + \frac{(A-B)^2 \rho}{4}
  + \mu^2 \rho
  - \lambda v^2\,\rho\,(2\rho^2+\chi^2)
  - \lambda' v^2\,\rho^3
  + \kappa v\,\rho\,\chi
  &=0, \label{SU3SO3EoMrho}\\
  \dv[2]{\chi}{r}
  + \frac{2}{r}\,\dv{\chi}{r}
  + \mu^2\,\chi
  - \lambda v^2\,\chi(2\rho^2+\chi^2)
  - \lambda' v^2\,\chi^3
  + \kappa v\,\rho^2
  &=0, \\
  \dv[2]{f}{r}
  + \frac{f-f^3}{r^2}
  + A^2 f
  - g^2v^2 f \rho^2
  &=0, \\
  \dv[2]{A}{r}
  + \frac{2}{r} \dv{A}{r}
  - \frac{2A f^2}{r^2}
  - g^2v^2 (A-B) \rho^2
  &=0, \\
  \dv[2]{B}{r}
  + \frac{2}{r} \dv{B}{r}
  + g'^2v^2 (A-B) \rho^2
  &=0.
\end{align}
\end{subequations}

Since the equations of motion \eqref{SU3SO3EoM} take the same form as those of the Cho--Maison monopole, the boundary conditions are imposed analogously. 
Requiring the regular behavior of $\mathrm{SU(3)_A}$ gauge field at the origin and finite energy at spatial infinity, we adopt
\begin{eqnarray}
  &\rho(0)=0,~~\chi(0)=\chi_0,~~f(0)=1,~~A(0)=0,~~B(0)=b_0, \nonumber\\
  &\rho(\infty)=1,~\chi(\infty)=1,~f(\infty)=0,~A(\infty)=A_0,~B(\infty)=B_0. \label{eq:boundary condition SU3 SO3}
\end{eqnarray}
Here, $\chi_0$, $b_0$, $A_0$, and $B_0$ are constants. 
In particular, since the equations for $f$, $A$, and $B$ have the same structure as in the Cho--Maison case, their boundary conditions follow in the same way.

\begin{figure}[ht]
  \centering
  \includegraphics[width=0.7\textwidth]{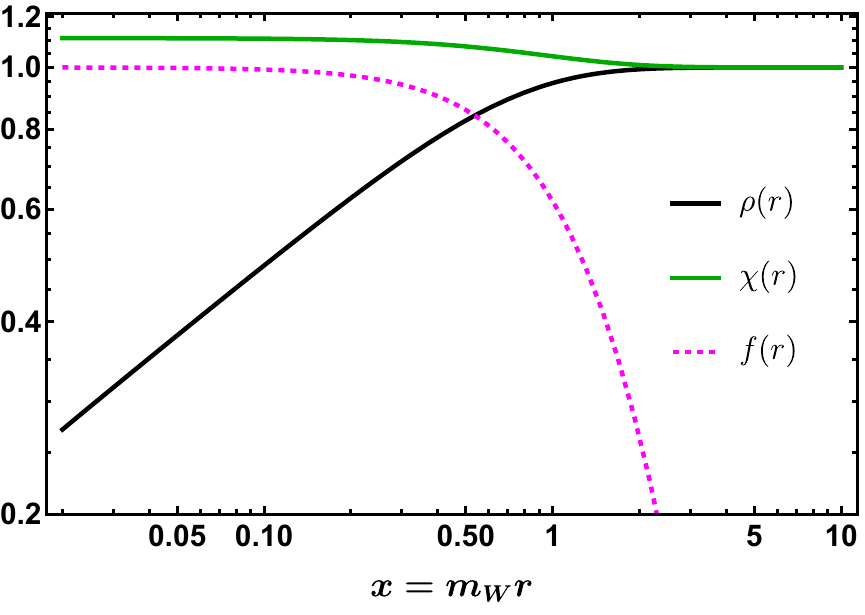}
  \caption{The profile functions of a generalized model \(\mathrm{SU(3)_A}\times\mathrm{SO(3)_B}\).
  We express the equations in dimensionless form by introducing the rescaled coordinate \(x = m_W r\), with \(m_W = gv\), and by normalizing the scalar fields by the vacuum expectation value \(v\).
  We work in units where \(m_W = 1\), and set \(g = g' = 1\), \(\lambda = \lambda' = 1\), \(\kappa = 0\), and \(\mu^2 = 2\). The black solid and magenta dashed curves correspond to \(\rho(x)\) and \(f(x)\), respectively.}
  \label{fig:SU3SO3_fit}
\end{figure}

Figure \ref{fig:SU3SO3_fit} shows an example of profile functions obtained from the equations of motion \eqref{SU3SO3EoM} with the boundary condition \eqref{eq:boundary condition SU3 SO3}. 
By setting $A=B=0$, we restrict ourselves to a purely magnetic sector, and hence the configuration represents a monopole solution.
Focusing on the scalar sector, the equation for $\rho$ in eq.~\eqref{SU3SO3EoMrho} around the origin behaves as
\begin{align}
  \dv[2]{\rho}{r}
  + \frac{2}{r} \dv{\rho}{r}
  - \frac{\rho}{2r^2}
  \simeq 0,
\end{align}
where we have used $f(r)\simeq 1$. 
This is the same equation as in the electroweak Cho--Maison case discussed in section~\ref{sec:electroweak monopole}, and thus leads to the same power-law behavior,
\begin{align}
  \rho(r) \propto r^{\tfrac{\sqrt{3}-1}{2}}.
\end{align}
This is in sharp contrast to the 't Hooft--Polyakov monopole, where $\rho(r)\propto r$.
The numerical solution shown in figure~\ref{fig:SU3SO3_fit} indeed exhibits this behavior, confirming that the present $\mathrm{SU(3)_A}\times\mathrm{SO(3)_B}$ model admits a Cho--Maison-like monopole configuration. 

In the closing of this section, we note that the construction of the monopole configuration in this section can be naturally generalized to a setup with gauge symmetry breaking $\mathrm{SU(N)} \times \mathrm{SO(N)} \to \mathrm{SO(N)_{diag}}$. Generalization of the Cho--Maison monopole explored in this section gives an implication for a general discussion on the condition of the appearance of Cho--Maison-like monopole.

\section{UV Completion by  't Hooft--Polyakov Monopole}\label{sec:UV completion}

As we have seen in the previous sections, the Cho--Maison monopoles suffer from its energy divergence at the spatial origin due to a singularity in the field configuration. Furthermore, its stability is even nontrivial. See refs.~\cite{Gervalle:2023electroweak, Gervalle:2022vxs, Mavromatos:2026nlp} for the previous discussions.
To address these problems, it is natural to consider the well-known 't Hooft--Polyakov monopole as a possible ultraviolet (UV) completion. 
Unlike the Cho--Maison monopole, the 't Hooft--Polyakov monopole has a smooth, nonsingular field configuration with finite energy, 
arising in non-Abelian gauge theories with spontaneous gauge symmetry breaking. 
Its stability is guaranteed by a nontrivial second homotopy group of the vacuum manifold $\pi_2(G/H)$.
This ensures that field configurations carrying a nonzero topological charge cannot be continuously deformed into the vacuum, 
thereby protecting the monopole solution against decay. 
In this sense, it serves as a physically well-motivated candidate to regularize the short-distance behavior of the electroweak monopole, 
thereby offering a consistent framework for studying its properties at high energies.

\subsection{Warm-up: $\mathrm{SU(2)_{A}}\times\mathrm{SU(2)_{B}}$ Gauge Theory} \label{sec:SU2SU2}
First, we demonstrate how the Cho-Maison monopole in SU(2) x U(1) can be regularized by the 't Hooft-Polyakov monopole by using the simplest toy model.
We consider a gauge theory with symmetry group 
$\mathrm{SU(2)_{A}} \times \mathrm{SU(2)_{B}}$ 
with two scalar fields \(\Phi\) and \(H\),
which transform under the gauge group as
\begin{align}
     \Phi \;\to\; U_{\mathrm{B}}\, \Phi\, U_{\mathrm{B}}^\dagger,\quad
     H \;\to\; U_{\mathrm{A}}\, H\, U_{\mathrm{B}}^\dagger,
\end{align}
with $U_{\mathrm{A}} \in \mathrm{SU(2)_{A}}$ 
and $U_{\mathrm{B}} \in \mathrm{SU(2)_{B}}$. 

The kinetic terms of the Lagrangian are given by
\begin{align}
\mathcal{L}_{\text{kin}}
  &=-\frac14 \, W^{a}_{\mathrm{A}\,\mu\nu} W^{a\,\mu\nu}_{\mathrm{A}}
    -\frac14 \, W^{a}_{\mathrm{B}\,\mu\nu} W^{a\,\mu\nu}_{\mathrm{B}} 
    + \mathrm{Tr}\!\left[(D_\mu H)^{\dagger} D^{\mu} H \right] 
    + \mathrm{Tr}\!\left[(D_\mu \Phi)(D^{\mu} \Phi)\right].
\end{align}
Here \(W^{a}_{\mathrm{A}\,\mu\nu}\) and \(W^{a}_{\mathrm{B}\,\mu\nu}\) denote the field strength tensors of the \(\mathrm{SU(2)_{A}}\) and \(\mathrm{SU(2)_{B}}\) gauge fields with gauge couplings \(g_{\mathrm{A}}\) and \(g_{\mathrm{B}}\), respectively.
The covariant derivatives of the scalar fields take the component form
\begin{align}
  D_\mu \Phi^{a}
  &= \partial_\mu \Phi^{a} + g_\mathrm{B}\,\varepsilon^{abc}\,W^{b}_{\mathrm{B}\mu}\,\Phi^{c}, \\[4pt]
  (D_\mu H)_{ij}
  &= \partial_\mu H_{ij}
   - \mathrm{i} g_\mathrm{A}(W^{a}_{\mathrm{A}\mu}T_{\mathrm{A}}^{a})_{ik} H_{kj}
   + \mathrm{i} g_\mathrm{B} H_{ik} (W^{a}_{\mathrm{B}\mu}T_{\mathrm{B}}^{a})_{kj}.
\end{align}
Here \(\varepsilon^{abc}\) denotes the totally antisymmetric tensor with \(\varepsilon^{123}=+1\), and the generators are taken as \(T^{a} = \sigma^a/2 \), normalized such that \(\mathrm{Tr}(T^{a}T^{b}) = \delta^{ab}/2 \).
The scalar potential is chosen as
\begin{align}
      V(\Phi,H)
  &=-\mu_{\Phi}^2 \Tr \Phi^2-\mu_{H}^2 \Tr H^{\dagger}H \nonumber \\
  &\quad + \lambda_1 \left(\Tr\Phi^2 \right)^2
  +\lambda_2 \left(\Tr  H^\dagger H \right)  ^2
  + \lambda_3 \, \Tr  H^\dagger H \, H^\dagger H.
  \label{SU2SU2_Potential}
\end{align}
This potential is bounded below if $\lambda_1 > 0$, $\lambda_2 + \lambda_3 > 0$, and $\lambda_2 + \lambda_3/2 > 0$ are satisfied.
For simplicity, we have omitted possible mixing terms $(\Tr\Phi^2)(\Tr HH^\dagger)$, $\Tr H \Phi H^\dagger$, and $\Tr H \Phi^2 H^\dagger$ in the scalar potential.

Following the analysis, we focus on the case with $\lambda_3 < 0$.\footnote{
If $\lambda_3 > 0$, the VEV of $H$ behaves as $\langle H \rangle \propto I$ and there is no nontrivial angular dependence of $\langle H \rangle$ around the 't Hooft--Polyakov monopole.}
The vacuum expectation value of the scalar fields as
\begin{align}
  \ev{\Phi} =
     \frac{v_{\Phi}}{2}
   \begin{pmatrix}
   1  & 0 \\
   0 & -1
   \end{pmatrix}, \quad
    \ev{H}
    =v_H
    \begin{pmatrix}
    0 & 0 \\
    0 & 1
    \end{pmatrix}, \label{eq:vev SU2SU2}
\end{align}
where
\begin{align}
   v_{\Phi} = \frac{\mu_{\Phi}}{\sqrt\lambda_1}, \quad
   v_{H} = \frac{\mu_{H}}{\sqrt{2\,(\lambda_{2}+\lambda_{3})}}.
\end{align}
Let us assume a hierarchy between the VEVs of the scalar field as $v_{\Phi} \gg v_H$. 
Then, we can interpret the pattern of spontaneous symmetry breaking is assumed to proceed in two steps,
\[
  \mathrm{SU(2)_{A}} \times \mathrm{SU(2)_{B}}
   \xrightarrow[]{\displaystyle \ev{\Phi}}
   \mathrm{SU(2)_{A}} \times \mathrm{U(1)_{B}}
   \xrightarrow[]{\displaystyle  \ev{H}}
   \mathrm{U}(1)_{\mathrm{diag}} \, .
\]
At the first stage of symmetry breaking, the vacuum manifold is
\[
   \frac{\mathrm{SU(2)_{B}}}{\mathrm{U(1)_B}} \;\simeq\; S^2,
\]
which carries a nontrivial second homotopy group $\pi_2(S^2)\cong \mathbb{Z}$.
While $H$ obtains a nonzero VEV around this monopole at the second stage of symmetry breaking, and this monopole remains stable because of the non-trivial second homotopy group $\pi_2( \mathrm{SU(2)_A} \times \mathrm{SU(2)_B} / \mathrm{U(1)_{diag}} )$.
Consequently, the theory admits 't~Hooft--Polyakov monopoles associated with this homotopy group, which are topologically stable.

At the first stage of gauge symmetry breaking, the gauge bosons associated with the broken generators of $\mathrm{SU(2)_B}\to\mathrm{U(1)_B}$ acquire masses of order $v_\Phi$. 
By integrating out these heavy degrees of freedom, we obtain an effective theory with $\mathrm{SU(2)_A} \times \mathrm{U(1)_B}$ gauge symmetry and a cutoff scale $\sim v_\Phi$.

In the following, we explicitly show the correspondence between the parameters of this effective theory and those of the ultraviolet theory.
The bi-doublet scalar field $H$ can be decomposed into two $\mathrm{SU(2)_L}$ doublets as
\begin{align}
  H = \begin{pmatrix}
    H_1 & H_2
  \end{pmatrix}.
\end{align}
Under $\mathrm{U(1)_B}$, the doublets $H_1$ and $H_2$ carry charges $-1/2$ and $+1/2$, respectively.
In terms of these fields, the scalar potential \eqref{SU2SU2_Potential} becomes
\begin{align}
    V_\mathrm{eff} (H_1, H_2) 
    &=-\mu^2_H (|H_1|^2 + |H_2|^2) \nonumber\\
    &\quad + \lambda_2 (|H_1|^2 + |H_2|^2)^2 + \lambda_3 (|H_1|^4 +2 (H_1^\dagger H_2) (H_2^\dagger H_1) + |H_2|^4),
    \label{eq:SU2SU2 effective potential}
\end{align}
where $|H_i|^2 \equiv H_i^\dagger H_i$.
The VEVs of $H_1$ and $H_2$ are given by
\begin{align}
    \langle H_1 \rangle = \left(\begin{array}{c} 0 \\ 0 \end{array}\right), \quad
    \langle H_2 \rangle = \left(\begin{array}{c} 0 \\ v_H \end{array}\right),
\end{align}
and the gauge symmetry is further broken to the diagonal subgroup
\begin{align}
  \mathrm{SU(2)_A} \times \mathrm{U(1)_B}
  \;\longrightarrow\;
  \mathrm{U(1)_{\mathrm{diag}}}.
\end{align}
The field $H_2$ contains three massless would-be NG bosons eaten by the massive gauge bosons, and one massive $\mathrm{U(1)_{diag}}$-neutral scalar corresponding to the Higgs boson. 
The field $H_1$ contains a massive $\mathrm{U(1)_{diag}}$-neutral complex scalar and a massless $\mathrm{U(1)_{diag}}$-charged complex scalar.

Let us discuss a monopole-like solution of the equation of motion. Since the VEV $\langle H_2\rangle$ does not induce mixing between $H_1$ and $H_2$ and no tachyonic direction of $H_1$ around $\langle H_1 \rangle = 0$, we can take $H_1 = 0$.

Substituting $H_1 = 0$ into the Lagrangian, we obtain the effective Lagrangian for the remaining doublet $H_2$,
\begin{align}
  \mathcal{L}_{\mathrm{eff}}
  \;=\;
  |D_\mu H_2|^2
  + \mu_H^2 |H_2|^2
  - (\lambda_2 + \lambda_3)\, |H_2|^4,
\end{align}
with the covariant derivative
\begin{align}
  D_\mu H_2
  \;=\;
  \Bigl(
    \partial_\mu
    - \mathrm{i}\, g_{\mathrm{A}}\, W^a_{\mathrm{A}\mu}\, T_{\mathrm{A}}^a
    - \mathrm{i}\, \frac{g_{\mathrm{B}}}{2} \, W^3_{\mathrm{B}\mu}
  \Bigr) H_2.
\end{align}
Comparing this Lagrangian with the electroweak Lagrangian reviewed in section~\ref{sec:electroweak monopole}, we can see that the scalar potential takes the same form as the Higgs potential.
We identify the gauge couplings and the parameters in the scalar potential as
\begin{align}
  g = g_{\mathrm{A}},  \quad
  g' = g_{\mathrm{B}}, \quad
  \mu^2 = \mu_H^2,  \quad
  \lambda = 2(\lambda_2 + \lambda_3). \label{eq:eft parameters}
\end{align}

Using this identification, the equation of motion for $H_2$ becomes identical to that of the Higgs field in the electroweak theory.

We now explicitly construct the 't Hooft--Polyakov monopole in the current setup. We take the scalar field configurations as
\begin{align}
    \Phi(x) 
    =\chi(r) U_2 \ev{\Phi} U_2^{\dagger}, \quad
       H(x) 
    =h(r) U_2 \ev{H} U_2^{\dagger},
\end{align}
where the vacuum expectation values are $v_{\Phi}=\mu_{\Phi}/\sqrt{\lambda_1},~v_{H}=\mu_{H}/\sqrt{2(\lambda_2+\lambda_3)} $.
Here $U_2(\theta,\varphi)$ is defined in eq.~\eqref{eq:u2}, and it rotates the scalar VEV as the hedgehog-like form.
For finite-energy configurations, the scalar fields are required to approach the VEV equivalent to eq.~\eqref{eq:vev SU2SU2} in the limit $r\to\infty$.
The gauge fields are assumed to take the same configuration as in the ’t Hooft--Polyakov monopole.
Therefore, the ansatz for the scalar and gauge fields can be summarized as
\begin{subequations}\label{SU2SU2Ansatz}
  \begin{align}
    \Phi(x)
    &=\frac{v_{\Phi}}{2}\,\chi(r)
    \begin{pmatrix}
      \cos\theta & e^{-\mathrm{i}\varphi}\sin\theta \\
      e^{\mathrm{i}\varphi}\sin\theta & -\cos\theta
    \end{pmatrix}
    =v_{\Phi} \chi(r)\frac{x^a}{r}\frac{\sigma^a}{2},\\
    H(x)
    &=\frac{v_{H}}{2}\,h(r)
    \begin{pmatrix}
      1-\cos\theta & -e^{-\mathrm{i}\varphi}\sin\theta \\
      -e^{\mathrm{i}\varphi}\sin\theta & 1+\cos\theta
    \end{pmatrix}
    =v_H h(r)\left(\frac{1}{2}1_2-\frac{x^a}{r}\frac{\sigma^a}{2} \right),\\
    W^{a}_{\mathrm{A} 0} &= W^{a}_{\mathrm{B} 0} = 0,\\
    W^{a}_{\mathrm{A} i}
    &=\frac{1}{g_{\mathrm{A}}}
    \left( 1-f_{\mathrm{A}}(r)\right)\,
    \varepsilon_{aij}\frac{x^{j}}{r^2},\quad W^{a}_{\mathrm{B} i}
    =\frac{1}{g_{\mathrm{B}}}\left(1-f_{\mathrm{B}}(r)\right)\,\varepsilon_{aij}\frac{x^{j}}{r^2},
  \end{align}
\end{subequations}
where the boundary conditions of $\chi(r)$, $h(r)$, $f_\mathrm{A}(r)$, and $f_\mathrm{B}(r)$ are
\begin{subequations}\label{SU2SU2BC}
  \begin{alignat}{5}
    \chi(0) &= 0, 
    &\quad h(0) &= 0, 
    &\quad f_{\mathrm{A}}(0) &= 1, 
    &\quad f_{\mathrm{B}}(0) &= 1, \\
    \chi(\infty) &= 1, 
    &\quad h(\infty) &= 1, 
    &\quad f_{\mathrm{A}}(\infty) &= 0, 
    &\quad f_{\mathrm{B}}(\infty) &= 0.
  \end{alignat}
\end{subequations} 
Substituting this ansatz into the equations of motion, we obtain the following reduced equations:

\begin{figure}[t]
  \centering
  \includegraphics[width=0.7\textwidth]{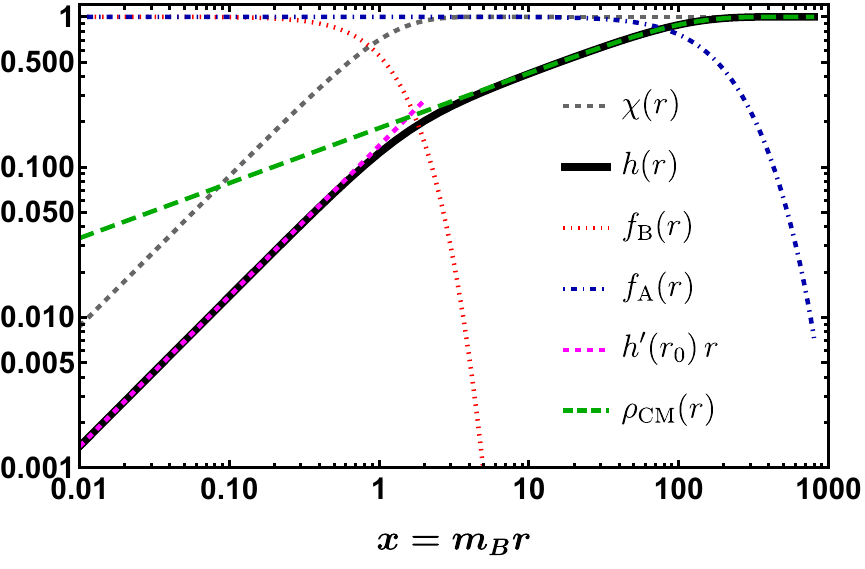}
  \caption{
    The profile functions of the \(\mathrm{SU(2)_A}\times\mathrm{SU(2)_B}\) toy model.
    We express the equations in dimensionless form by introducing the rescaled coordinate \(x = m_B r\), with \(m_B^2 = g_{\mathrm{B}}^2 v_\Phi^2\).
    We work in units where \(m_B = 1\), and set \(g_{\mathrm{A}}= g_{\mathrm{B}} = 1\), \(v_\Phi = 100 v_H\), and \(v_H = 0.01\).
    The gray dashed, black solid, red dotted, and blue dot-dashed curves correspond to \(\chi(x)\), \(h(x)\), \(f_{\mathrm{B}}(x)\), and \(f_{\mathrm{A}}(x)\), respectively.
    The magenta dashed curve shows the linear behavior \(h'(0)\,x\), representing the near-origin asymptotic form of the ’t~Hooft--Polyakov monopole, while the green dashed curve denotes the Cho--Maison monopole profile \(\rho_{\mathrm{CM}}(x)\), whose detail is explained in the main text.}
  \label{fig:SU2SU2Fig}
\end{figure}
\begin{subequations}\label{SU2SU2EoM}
\begin{align}
  \dv[2]{\chi}{r}
  + \frac{2}{r}\, \dv{\chi}{r}
  - \frac{2 f_{\mathrm{B}}(r)^{2}}{r^{2}}\, \chi(r)
  - \mu_{\Phi}^{2}\, \bigl(\chi(r)^{2}-1\bigr)\, \chi(r)
  &=0,\\
  \dv[2]{h}{r}
  + \frac{2}{r}\, \dv{h}{r}
  - \frac{\bigl(f_{\mathrm{A}}(r)+f_{\mathrm{B}}(r)\bigr)^{2}}{2\, r^{2}}\, h(r)
  - \mu_{H}^{2}\, \bigl(h(r)^{2}-1\bigr)\, h(r)
  &=0,\\
  \dv[2]{f_{\mathrm{B}}}{r}
  - \frac{f_{\mathrm{B}}(r)}{r^{2}}\, \bigl(f_{\mathrm{B}}(r)^{2}-1\bigr)
  - g_{\mathrm{B}}^{2}\, f_{\mathrm{B}}(r)
  \left(v_{\Phi}^{2}\, \chi(r)^{2}
  + \frac{1}{2}\, v_{H}^{2}\, h(r)^{2} \right)
  &=0,\\
  \dv[2]{f_{\mathrm{A}}}{r}
  - \frac{f_{\mathrm{A}}(r)}{r^{2}}\, \bigl(f_{\mathrm{A}}(r)^{2}-1\bigr)
  - \frac{1}{2}\, g_{\mathrm{A}}^{2}\, v_{H}^{2}\,
  f_{\mathrm{A}}(r)\, h(r)^{2}
  &=0.
\end{align}
\end{subequations}
Figure \ref{fig:SU2SU2Fig} shows an example of solution of eq.~\eqref{SU2SU2EoM} with the boundary condition \eqref{SU2SU2BC}. 
The key point illustrated by figure~\ref{fig:SU2SU2Fig} is that the scalar field \(H\), which a Cho--Maison-like behavior in the low-energy effective theory, is smoothly completed into an ’t~Hooft--Polyakov-like configuration in the ultraviolet regime.
To see this, let us examine the profile functions in detail.

Because of the two-step symmetry breaking structure of the model, the behavior of this monopole solution is characterized by two length scales, $r_1$ and $r_2$, which are the inverse of the mass of massive Higgs boson responsible for each step of symmetry breaking:
\begin{align}
    r_1 \sim \frac{1}{\sqrt{\lambda_1}v_{\Phi}}, \quad r_2 \sim \frac{1}{\sqrt{\lambda_2+\lambda_3}v_{H}}.
\end{align}
In the inner region, $r \lesssim r_1$, we can see that both of the scalar profile function $\chi(r)$ and $h(r)$ behave as $\propto r$ and $f_A \simeq f_B \simeq 1$. This behavior reflects the boundary condition of the 't Hooft--Polyakov monopole we have imposed.
At $r\sim r_1$, $\chi(r)$ reaches $\simeq 1$ and the VEV of $\Phi$ breaks $\mathrm{SU(2)_B}$ gauge symmetry to $\mathrm{U(1)_B}$.

In the intermediate region, \(r_1 \lesssim r \lesssim r_2\), the gauge symmetry is effectively reduced to \(\mathrm{SU(2)_A}\times \mathrm{U(1)_B}\) due to \(\chi(r)\simeq 1\), and the effective field theory with the scalar potential given in eq.~\eqref{eq:SU2SU2 effective potential} provides an appropriate description. 
Since the symmetry breaking pattern
\(  
\mathrm{SU(2)_A} \times \mathrm{U(1)_B}
\to
\mathrm{U(1)_{diag}}
\)
does not admit the 't~Hooft--Polyakov monopole, the profile functions in the intermediate region deviate from those of the 't~Hooft--Polyakov monopole and instead exhibit Cho--Maison-like behavior. 

In figure~\ref{fig:SU2SU2Fig}, we compare \(h(r)\) with the Cho--Maison monopole profile function \(\rho_{\mathrm{CM}}(r)\), obtained as a solution to the equations of motion~\eqref{eq:eom CM monopole} with the parameter identification~\eqref{eq:eft parameters}. We find remarkable agreement between \(h(r)\) and \(\rho_{\mathrm{CM}}(r)\) for \(r \gtrsim r_1\). 
At \(r \simeq r_2\), \(h(r)\) also approaches unity. 
Consequently, in the asymptotic region \(r>r_2\), where \(H\) acquires a vacuum expectation value, the gauge symmetry is further broken to the diagonal subgroup \(\mathrm{U(1)}_{\mathrm{d}}\).

To summarize, at $r \gtrsim r_1$, the 't Hooft--Polyakov monopole in the current model behaves as the Cho--Maison monopole from $\mathrm{SU(2)_A} \times \mathrm{U(1)_B} \to \mathrm{U(1)_{diag}}$. Figure \ref{fig:SU2SU2Fig} shows a smooth transition from the 't Hooft--Polyakov monopole to the Cho--Maison monopole. In this sense, the 't Hooft--Polyakov monopole we have discussed can be regarded a UV completion of the Cho--Maison monopole from $\mathrm{SU(2)_A} \times \mathrm{U(1)_B} \to \mathrm{U(1)_{diag}}$.
Note that the model discussed so far should be regarded as a toy model, and a more realistic candidate of UV completion of the Cho--Maison electroweak monopole \cite{Cho:1996qd} based on a GUT model will be discussed in the next subsection.

\subsection{GUT Example: Pati--Salam Model} \label{sec:PatiSalam}
The Pati--Salam model \cite{Pati:1974yy}, based on the gauge group 
$\mathrm{SU(4)_{C}}\times \mathrm{SU(2)_{L}}\times \mathrm{SU(2)_{R}}$,
provides a well-known example of a semi-simple grand unified theory. 
The extended color group $\mathrm{SU(4)_{C}}$ unifies quarks and leptons, 
while the left-right symmetry encoded in $\mathrm{SU(2)_{L}}\times \mathrm{SU(2)_{R}}$ 
offers a natural embedding of the electroweak sector. 
Here we will show that a 't Hooft--Polyakov monopole in this model can be regarded as the electroweak Cho--Maison monopole in the low energy effective description.\footnote{
Note that, in the monopole of the minimal $\mathrm{SU(5)}$ GUT \cite{Georgi:1974sy}, the Higgs doublet field obtains its nonzero VEV at the origin \cite{Eckert:1983ze, Eckert:1983bq} and it does not behave as the Cho--Masion monopole.}

In this paper, we focus on a minimal version of Pati--Salam model with  two scalar fields $\Phi$ and $H$ with transformation rules
\begin{align}
     \Phi &\;\to\; U_\mathrm{C}\, \Phi\, U_\mathrm{R}^\dagger,
    \quad
       H \;\to\; U_{\mathrm{L}}\, H\, U_\mathrm{R}^\dagger,
\end{align}
where $U_{\mathrm{C}} \in \mathrm{SU(4)_{C}}$, $U_{\mathrm{L}} \in \mathrm{SU(2)_{L}}$, and $U_\mathrm{R} \in \mathrm{SU(2)_{R}}$. 
The kinetic terms of the Lagrangian are given by
\begin{align}
\mathcal{L}_{\text{kin}}
  &=-\frac{1}{4}G^a_{\mu\nu}G^{a \mu\nu} -\frac14 \, W^{a}_{\mathrm{L}\,\mu\nu} W^{a\,\mu\nu}_{\mathrm{L}}
    -\frac14 \, W^{a}_{\mathrm{R}\,\mu\nu} W^{a\,\mu\nu}_{\mathrm{R}} \nonumber \\[4pt]
  &\quad
    + \mathrm{Tr}\!\left[(D_\mu H)^{\dagger} D^{\mu} H \right] 
    + \mathrm{Tr}\!\left[(D_\mu \Phi)^{\dagger} (D^{\mu} \Phi)\right],
\end{align}
where the covariant derivative of scalar fields are defined as
\begin{align}
    D_\mu \Phi = \partial_\mu \Phi - \i g_{\mathrm{C}} G_{\mu} \Phi + \i g_{\mathrm{R}} \Phi W_{{\mathrm{R}}\mu}, \quad
    D_\mu H = \partial_\mu H - \i g_{\mathrm{L}} W_{{\mathrm{L}}\mu} H + \i g_{\mathrm{R}} H W_{{\mathrm{R}}\mu}.
\end{align}
$G_\mu$, $W_{L\mu}$, and $W_{R\mu}$ are matrix form of gauge fields for $\mathrm{SU(4)_C}$, $\mathrm{SU(2)_L}$, and $\mathrm{SU(2)_R}$ gauge symmetry, respectively.
Here \(G^{a}_{\mu\nu}\) is the field strength tensor of the \(\mathrm{SU(4)_{C}}\) gauge fields \(G^{a}_{\mu}\) with gauge coupling \(g_4\), and \(W^{a}_{\mathrm{L}\,\mu\nu}\) and \(W^{a}_{\mathrm{R}\,\mu\nu}\) are those of \(\mathrm{SU(2)_{\mathrm{L}}}\) and \(\mathrm{SU(2)_{\mathrm{R}}}\) with gauge couplings \(g_{\mathrm{L}}\) and \(g_{\mathrm{R}}\), respectively.
The scalar potential is chosen as
\begin{align}
        V(\Phi,H)
  &=-\mu_{\Phi}^2 \Tr \Phi^{\dagger} \Phi-\mu_{H}^2 \Tr H^{\dagger}H \nonumber  \\
  & \quad +\lambda_1 \,  \left(\Tr  \Phi^\dagger \Phi \right)  ^2
+ \lambda_2 \, \Tr  \Phi^\dagger \Phi \, \Phi^\dagger \Phi
  +\lambda_3 \,  \left(\Tr  H^\dagger H \right)  ^2
+ \lambda_4 \, \Tr  H^\dagger H \, H^\dagger H.  
\end{align}
This potential is bounded below if $\lambda_1 + \lambda_2 > 0$, $\lambda_1 + \lambda_2/2 > 0$, $\lambda_3 + \lambda_4 > 0$, and $\lambda_3 + \lambda_4/2 > 0$ are satisfied. 
As in the toy model discussed in the previous section \ref{sec:SU2SU2}, we have, for simplicity, omitted possible mixing terms $\Tr \Phi^\dagger \Phi H^\dagger H $ and $(\Tr \Phi^\dagger \Phi)(\Tr H^\dagger H)$ in the scalar potential.  
Such a term can be included in general, but it does not play an essential role in the construction of the monopole solution and therefore does not affect the qualitative discussion presented here.  

Assuming $\lambda_2 < 0$ and $\lambda_4 < 0$, one finds that the vacuum expectation values of the scalar fields are given by
\begin{align}
  \ev{\Phi} =
    \begin{pmatrix}
    0 & 0 \\
    0 & 0 \\
    0 & 0 \\
    0 & v_{\Phi}
   \end{pmatrix},\quad
    \ev{H}
    =
    \begin{pmatrix}
    0 & 0\\
    0 & v_H
    \end{pmatrix}.
\end{align}
where
\begin{align}
    v_\Phi = \frac{\mu_\Phi}{ \sqrt{ 2\,(\lambda_1+\lambda_2)} } , \quad
    v_H = \frac{\mu_{H}}{ \sqrt{ 2\,(\lambda_{3}+\lambda_{4})} }.
\end{align}

Assuming the hierarchy between VEV as $v_\Phi \gg v_H$, we can interpret the symmetry breaking pattern as two steps:
\begin{align}
  \mathrm{SU(4)_{C}} \times \mathrm{SU(2)_{L}} \times \mathrm{SU(2)_{R}}
  & \xrightarrow[]{\displaystyle \ev{\Phi}}
  \mathrm{SU(3)_{C}} \times \mathrm{SU(2)_{L}} \times \mathrm{U(1)_{Y}}\\
  & \xrightarrow[]{\displaystyle \ev{H}}
  \mathrm{SU(3)_{C}} \times \mathrm{U(1)_{\mathrm{EM}}} \, .
\end{align}
At the first stage $\mathrm{U(1)_{Y}}$ is the unbroken linear combination inside
$\mathrm{SU(4)_{C}}\times \mathrm{SU(2)_{R}}$ (canonically, $Y = T^3_{\mathrm{R}}+\tfrac12(B-L)$).
Here $\mathrm{SU(2)_{L}}$ remains unbroken and thus plays no role in the topology of the vacuum manifold.
So the vacuum manifold is
\[
  \mathcal{M}\simeq
  \frac{\mathrm{SU(4)_{C}}\times \mathrm{SU(2)_{R}}}
       {\mathrm{SU(3)_{C}}\times \mathrm{U(1)_{Y}}}\,.
\]
Using that $\pi_{1}(\mathrm{SU}(n))=0\,(n \ge2)$ and $\pi_{1}(\mathrm{U}(1))\cong\mathbb{Z}$, and that
$G=\mathrm{SU(4)_{C}}\times \mathrm{SU(2)_{R}}$ is simply connected, the long exact
sequence in homotopy gives
\[
  \pi_{2}(\mathcal{M})\;\cong\;\pi_{1}\!\big(\mathrm{SU(3)_{C}}\times \mathrm{U(1)_{Y}}\big)
  \;\cong\;\mathbb{Z}.
\]
Similar as section \ref{sec:SU2SU2},
while $H$ obtains a nonzero VEV around this monopole at the second stage of symmetry breaking, and this monopole remains stable because of the non-trivial second homotopy group $\pi_2( \mathrm{SU(4)_{C}} \times \mathrm{SU(2)_{L}} \times \mathrm{SU(2)_{R}} / \mathrm{SU(3)_{C}} \times \mathrm{U(1)_{\mathrm{EM}}} )$.
Therefore the breaking admits topologically stable ’t~Hooft--Polyakov monopoles.

As in section~\ref{sec:SU2SU2}, since the symmetry breaking occurs at the scale \(v_\Phi\), the gauge bosons associated with the broken generators of 
\(\mathrm{SU(4)_C} \times \mathrm{SU(2)_L} \times \mathrm{SU(2)_R} \to \mathrm{SU(3)_C} \times \mathrm{SU(2)_L} \times \mathrm{U(1)_{Y}}\) 
acquire masses of order \(v_\Phi\).
By integrating out heavy degrees of freedom, we obtain an effective theory with \(\mathrm{SU(3)_C} \times \mathrm{SU(2)_L} \times \mathrm{U(1)_{Y}}\) gauge symmetry with a cutoff scale $\sim v_\Phi$.
In the following, we make explicit the correspondence between the parameters of this effective theory and those of the ultraviolet theory.
The bi-doublet scalar field $H$ can be decomposed into two $\mathrm{SU(2)_L}$ doublets as
\begin{align}
  H = \begin{pmatrix}
    H_1 & H_2
  \end{pmatrix}.
\end{align}
Under $\mathrm{U(1)_R}$, the doublets $H_1$ and $H_2$ carry charges $-1/2$ and $+1/2$, respectively.
Since the hypercharge is given by
\begin{align}
  Y = T^3_{\mathrm{R}} + \frac{B-L}{2},
\end{align}
and the bi-doublet $H$ is neutral under $\mathrm{U(1)_{B-L}}$, it follows that $Y = T^3_{\mathrm{R}}$ for its components.
Therefore, $H_2$ carries hypercharge $Y=+1/2$, and can be identified with the Standard Model Higgs doublet.
In terms of these fields, the scalar potential \eqref{SU2SU2_Potential} becomes
\begin{align}
    V_\mathrm{eff} (H_1, H_2) 
    &=-\mu^2_H (|H_1|^2 + |H_2|^2) \nonumber\\
    &\quad + \lambda_3 (|H_1|^2 + |H_2|^2)^2 + \lambda_4 (|H_1|^4 +2 (H_1^\dagger H_2) (H_2^\dagger H_1) + |H_2|^4)
    \label{eq:PS effective potential}
\end{align}
where $|H_i|^2 \equiv H_i^\dagger H_i$.
After the second stage of symmetry breaking, the gauge symmetry is further broken to the diagonal subgroup
\begin{align}
 \mathrm{SU(3)_C} \times \mathrm{SU(2)_L} \times \mathrm{U(1)_{Y}} \to \mathrm{SU(3)_C} \times \mathrm{U(1)_{EM}}.
\end{align}
$H_2$ includes massless would-be NG bosons eaten by massive gauge bosons, and a massive $\mathrm{U(1)_{EM}}$ neutral boson corresponding to the Higgs boson. $H_1$ includes a massive $\mathrm{U(1)_{EM}}$ neutral complex scalar and a massless $\mathrm{U(1)_{EM}}$ charged complex scalar boson.

Let us discuss a monopole-like solution of the equation of motion. Since there is no mixing between $H_1$ and $H_2$ and no tachyonic direction of $H_1$ around $\langle H_1 \rangle = 0$, we can take $H_1 = 0$.

Substituting this into the Lagrangian, we obtain the effective theory for the remaining light doublet $H_2$,
\begin{align}
  \mathcal{L}_{\mathrm{eff}}
  \;=\;
  |D_\mu H_2|^2
  + \mu_H^2 |H_2|^2
  - (\lambda_3 + \lambda_4)\, |H_2|^4,
\end{align}
with the covariant derivative
\begin{align}
  D_\mu H_2
  \;=\;
  \Bigl(
    \partial_\mu
    - \mathrm{i}\, g_{\mathrm{L}}\, W^a_{\mathrm{L}\mu}\, T_{\mathrm{L}}^a
    - \mathrm{i}\, \frac{g_{\mathrm{R}}}{2} \, W^3_{\mathrm{R}\mu}
  \Bigr) H_2.
\end{align}
Comparing this Lagrangian with the electroweak Lagrangian reviewed in section~\ref{sec:electroweak monopole}, we can see that the scalar potential takes the same form as the Higgs potential.
We identify the gauge couplings and the parameters in the scalar potential as
\begin{align}
  g = g_{\mathrm{A}},  \quad
  g' = g_{\mathrm{B}}, \quad
  \mu^2 = \mu_H^2,  \quad
  \lambda = 2(\lambda_2 + \lambda_3). \label{eq:eft parameters patisalam}
\end{align}

Using this identification, the equation of motion for $H_2$ becomes identical to that of the Higgs field in the electroweak theory.

After the electroweak breaking driven by $\ev{H}$, the ’t~Hooft--Polyakov monopoles present at the first stage reduce to objects that resemble the electroweak (Cho--Maison) monopole from the low-energy viewpoint. 
Thus, just as in the warm-up model, the Pati--Salam theory naturally accommodates monopole solutions that interpolate between the ultraviolet ’t~Hooft--Polyakov monopoles and the infrared electroweak monopoles. 
The technical analysis of their profiles and stability can be carried out in essentially the same manner as in the simpler $\mathrm{SU(2)_{L}}\times \mathrm{SU(2)_{R}}$ case.

We now explicitly construct the 't Hooft--Polyakov monopole in the current setup.
The hedgehog ansatz can be constructed in precisely the same manner as in the toy model. We take the ansatz for $\Phi$ and $H$ as
\begin{align}
  \Phi(x)
  &~=~\chi(r)\, U_{4}\ev{\Phi}U_{2}^{\dagger}
  ~=~\frac{v_{\Phi}}{2}\, \chi(r)
  \begin{pmatrix}
    0 & 0 \\
    0 & 0 \\
    \cos\theta & e^{-\mathrm{i}\varphi}\sin\theta \\
    e^{\mathrm{i}\varphi}\sin\theta & -\cos\theta
  \end{pmatrix},\label{eq:ansatz_Phi}
  \\[1ex]
  H(x)
  &~=~h(r)\, U_{2}\ev{H}U_{2}^{\dagger}
  ~=~\frac{v_{H}}{2}\, h(r)
  \begin{pmatrix}
    1+\cos\theta & e^{-\mathrm{i}\varphi}\sin\theta \\
    e^{\mathrm{i}\varphi}\sin\theta & 1-\cos\theta
  \end{pmatrix},\label{eq:ansatz_H}
\end{align}
where $U_2(\theta,\varphi)$ defined in eq.~\eqref{eq:u2},
and $U_4(\theta,\varphi)$ is defined as
\begin{align}
    U_4(\theta,\varphi) = \begin{pmatrix}
        1 & & \\
        & 1 & \\
        & & U_2(\theta,\varphi)
    \end{pmatrix}.
\end{align}
 
For gauge fields, we take
\begin{align}
    G_0 &= 0, \quad W_{L0} = W_{R0} = 0, \\
  G_i &= \frac{1-f_4(r)}{g_4} \begin{pmatrix}
    0 && \\
    & 0 & \\
    && \varepsilon_{aij} \dfrac{x^j}{r^2} \dfrac{\sigma^a}{2}
  \end{pmatrix}, \label{eq:ansatz_G}\\    
  W_{\mathrm{L} i}
  &=
  \frac{1-f_{\mathrm{L}}(r)}{g_{\mathrm{L}}}\,\varepsilon_{aij}\,\frac{x^{j}}{r^{2}} \frac{\sigma^a}{2},\\
  W_{\mathrm{R} i}
  &=\frac{1-f_{\mathrm{R}}(r)}{g_{\mathrm{R}}}\,\varepsilon_{aij}\,\frac{x^{j}}{r^{2}} \frac{\sigma^a}{2}.\label{eq:ansatz_W}
\end{align}

The boundary conditions of $\chi(r)$, $h(r)$, $f_4(r)$, $f_\mathrm{L}(r)$, and $f_\mathrm{R}(r)$ are
\begin{subequations}\label{PSBC}
  \begin{alignat}{5}
    \chi(0) &= 0, 
    &\quad h(0) &= 0, 
    &\quad f_{4}(0) &= 1, 
    &\quad f_{\mathrm{L}}(0) &= 1, 
    &\quad f_{\mathrm{R}}(0) &= 1, \\
    \chi(\infty) &= 1, 
    &\quad h(\infty) &= 1, 
    &\quad f_{4}(\infty) &= 0, 
    &\quad f_{\mathrm{L}}(\infty) &= 0, 
    &\quad f_{\mathrm{R}}(\infty) &= 0.
  \end{alignat}
\end{subequations}

\begin{figure}[t]
  \centering
  \includegraphics[width=0.7\textwidth]{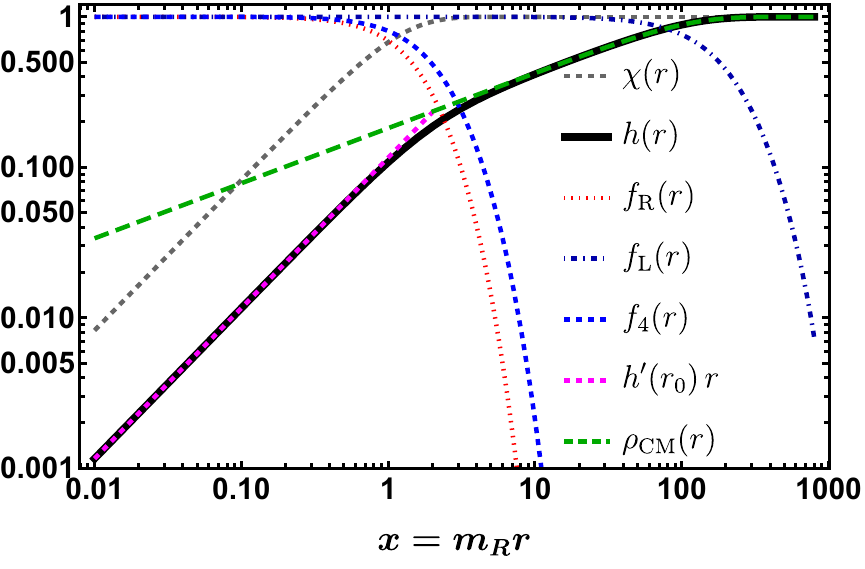}
  \caption{The profile functions of the \(\mathrm{SU(4)_C}\times\mathrm{SU(2)_L}\times\mathrm{SU(2)_R}\) toy model.
  We express the equations in dimensionless form by introducing the rescaled coordinate \(x = m_R r\), with \(m_R = g_R v_\Phi/\sqrt{2}\).
  We work in units where \(m_R = 1\), and set \(g_4 = g_R = g_L = 1\), \(v_\Phi = 100 v_H\), and \(v_H = 0.01\).
  The gray dashed, black solid, red dotted, blue dot-dashed, and blue dashed curves correspond to \(\chi(x)\), \(h(x)\), \(f_{\mathrm{R}}(x)\), \(f_{\mathrm{L}}(x)\), and \(f_{4}(x)\), respectively.The magenta dashed curve shows the linear behavior \(h'(0)\,x\), representing the near-origin asymptotic behavior of the ’t~Hooft--Polyakov monopole, while the green dashed curve denotes the Cho--Maison monopole profile \(\rho_{\mathrm{CM}}(x)\), whose detail is explained in the main text.}
  \label{fig:PatiFig}
\end{figure}

We then obtain four series of differential equations.

\begin{subequations}\label{PSEoM}
\begin{align}
  \dv[2]{\chi}{r}
  + \frac{2}{r}\, \dv{\chi}{r}
  - \frac{\bigl(f_{4}(r)+f_{\mathrm{R}}(r)\bigr)^{2}}{2\, r^{2}}\, \chi(r)
  - \mu_{\Phi}^{2}\, \bigl(\chi(r)^{2}-1\bigr)\, \chi(r)
  &=0,\\
  \dv[2]{h}{r}
  + \frac{2}{r}\, \dv{h}{r}
  - \frac{\bigl(f_{\mathrm{L}}(r)+f_{\mathrm{R}}(r)\bigr)^{2}}{2\, r^{2}}\, h(r)
  - \mu_{H}^{2}\, \bigl(h(r)^{2}-1\bigr)\, h(r)
  &=0,\\
  \dv[2]{f_{4}}{r}
  - \frac{f_{4}(r)}{r^{2}}\, \bigl(f_{4}(r)^{2}-1\bigr)
  - \frac{1}{2}\, g_{4}^{2}\, v_{\Phi}^{2}\, f_{4}(r)\, \chi(r)^{2}
  &=0,\\
  \dv[2]{f_{\mathrm{R}}}{r}
  - \frac{f_{\mathrm{R}}(r)}{r^{2}}\, \bigl(f_{\mathrm{R}}(r)^{2}-1\bigr)
  - \frac{1}{2}\, g_{\mathrm{R}}^{2}\, f_{\mathrm{R}}(r)
  \left(v_{\Phi}^{2}\, \chi(r)^{2}+ v_{H}^{2}\, h(r)^{2}\right)
  &=0,\\
  \dv[2]{f_{\mathrm{L}}}{r}
  - \frac{f_{\mathrm{L}}(r)}{r^{2}}\, \bigl(f_{\mathrm{L}}(r)^{2}-1\bigr)
  - \frac{1}{2}\, g_{\mathrm{L}}^{2}\, v_{H}^{2}\,
  f_{\mathrm{L}}(r)\, h(r)^{2}
  &=0.
\end{align}
\end{subequations}
The resulting profiles are shown in figure~\ref{fig:PatiFig}. Similar to the 't Hooft--Polyakov monopole discussed in section \ref{sec:SU2SU2}, the scalar field \(H\), which behaves as a Cho--Maison monopole in the low-energy effective theory, is smoothly completed into a regular ’t~Hooft--Polyakov-like configuration in the ultraviolet. Two length scales characterize the monopole profile:
\begin{align}
    r_1 \sim \frac{1}{\sqrt{2(\lambda_1+\lambda_2)}v_\Phi} ,\quad
    r_2 \sim \frac{1}{\sqrt{2(\lambda_3+\lambda_4)}v_H}.
\end{align}
$r_1$ is the inverse of the Higgs boson mass for symmetry breaking $\mathrm{SU(4)_C} \times \mathrm{SU(2)_L} \times \mathrm{SU(2)_R} \to \mathrm{SU(3)_C} \times \mathrm{SU(2)_L} \times \mathrm{U(1)_Y}$, and $r_2$ is the inverse of the electroweak scale.
In the inner region, $r \lesssim r_1$, the scalar profile function $\chi(r)$ and $h(r)$ behave as $\propto r$, and this behavior reflects the regular behavior and the boundary condition of 't Hooft--Polyakov monopole.
At $r\sim r_1$, $\chi(r)$ reaches $\simeq 1$ and the VEV of $\Phi$ breaks the gauge symmetry $\mathrm{SU(4)_C} \times \mathrm{SU(2)_L} \times \mathrm{SU(2)_R}$.
In the intermediate region, $r_1 \lesssim r \lesssim r_2$, an effective field theory with the SM gauge group $\mathrm{SU(3)_C} \times \mathrm{SU(2)_L} \times \mathrm{U(1)_Y}$ and the scalar potential given in eq.~\eqref{eq:PS effective potential} provides an effective description.
The behavior of the profile functions in the intermediate region deviates from that of the 't Hooft--Polyakov monopole around the origin, and agrees with the profile function of the Cho--Maison electroweak monopole.
In figure \ref{fig:PatiFig}, we compare $h(r)$ with the Cho--Maison monopole profile function $\rho_\mathrm{CM}(r)$, which is obtained as the solution of the equations of motion \eqref{eq:eom CM monopole} with the parameter identification \eqref{eq:eft parameters patisalam}. 
At $r \simeq r_2$, $h(r)$ also approaches unity.
Thus, in the asymptotic region \(r > r_2\), the gauge symmetry is further broken down to $\mathrm{U(1)}_{\mathrm{EM}}$ by the VEV of $H$.

Therefore, as we expected, the 't Hooft--Polyakov monopole in a minimal Pati--Salam model behaves as the Cho--Maison electroweak monopole.
Figure \ref{fig:PatiFig} shows a smooth transition from the 't Hooft--Polyakov monopole to the Cho--Maison monopole, and the monopole in the Pati--Salam model can be regarded a realistic UV completion of the Cho--Maison electroweak monopole discussed in ref.~\cite{Cho:1996qd}.

\subsection{Scope of the UV Completion for Generalized Cho--Maison Monopoles}
\label{sec:UVscope}

Let us comment on how the UV completion mechanism discussed above extends to
the generalized Cho--Maison monopoles introduced in section~\ref{sec:generalization}.
The construction does not rely on a special property of the electroweak
embedding itself.  The essential point is that the gauge group which gives the
singular low-energy gauge field should arise as an unbroken subgroup of a
larger simply connected non-Abelian gauge group.  Then the singular core of the
low-energy Cho--Maison-type configuration can be replaced by a regular
't~Hooft--Polyakov core associated with the first stage of symmetry breaking.

For example, the \(\mathrm{SU(3)_A}\times \mathrm{SO(3)_B}\) model discussed in
section~\ref{sec:generalization} can be embedded into a theory with gauge group
\begin{align}
  \mathrm{SU(3)_A}\times \mathrm{SU(3)_B}.
\end{align}
Introduce a scalar field \(\Sigma\) in the symmetric representation\footnote{
The symmetric representation appears in the decomposition
\(\mathbf{3}\otimes\mathbf{3}=\mathbf{6}\oplus\overline{\mathbf{3}}\).
Choosing the symmetric tensor allows us to represent \(\Sigma\) as a complex
symmetric \(3\times3\) matrix transforming as
\(\Sigma\to U_{\rm B}\Sigma U_{\rm B}^{T}\).  Then the VEV
\(\langle\Sigma\rangle\propto I_3\) preserves the subgroup satisfying
\(U_{\rm B}U_{\rm B}^{T}=I_3\), namely \(\mathrm{SO(3)_B}\).
}
of
\(\mathrm{SU(3)_B}\),
\begin{align}
  \Sigma \sim (\mathbf{1},\mathbf{6}),\quad
  \Sigma \to U_{\rm B}\Sigma U_{\rm B}^{T},
\end{align}
and a bifundamental scalar field
\begin{align}
  \Phi \sim (\mathbf{3},\overline{\mathbf{3}}),\quad
  \Phi \to U_{\rm A}\Phi U_{\rm B}^{\dagger}.
\end{align}
If \(\Sigma\) develops a VEV proportional to the identity matrix,
\begin{align}
  \langle \Sigma\rangle = v_{\Sigma} I_3,
\end{align}
then \(\mathrm{SU(3)_B}\) is broken to the subgroup satisfying
\(U_{\rm B}U_{\rm B}^{T}=I_3\), namely \(\mathrm{SO(3)_B}\).  At energies below
the scale \(v_{\Sigma}\), the field \(\Phi\) is precisely the scalar field used
in the \(\mathrm{SU(3)_A}\times\mathrm{SO(3)_B}\) model, since the
\(\overline{\mathbf{3}}\) of \(\mathrm{SU(3)_B}\) restricts to the vector
representation of \(\mathrm{SO(3)_B}\).  Taking the VEV of $\Phi$ as given in \eqref{SU3SO3VEVPhi}, 
the overall symmetry breaking pattern proceeds in two steps:
\begin{align}
  \mathrm{SU(3)_A}\times\mathrm{SU(3)_B}
  \xrightarrow[]{\displaystyle \langle\Sigma\rangle}
  \mathrm{SU(3)_A}\times\mathrm{SO(3)_B}
  \xrightarrow[]{\displaystyle \langle\Phi\rangle}
  \mathrm{SO(3)_{\rm diag}} .
\end{align}
The first vacuum manifold is
\begin{align}
  \frac{\mathrm{SU(3)_B}}{\mathrm{SO(3)_B}},
\end{align}
and, because \(\mathrm{SU(3)_B}\) is simply connected,
\begin{align}
  \pi_2\!\left(\frac{\mathrm{SU(3)_B}}{\mathrm{SO(3)_B}}\right)
  \simeq \pi_1(\mathrm{SO(3)_B})
  \simeq \mathbb Z_2 .
\end{align}
Thus the ultraviolet theory admits a regular 't~Hooft--Polyakov monopole with
\(\mathbb Z_2\) charge.  In the hierarchy \(v_{\Sigma}\gg v_{\Phi}\), the heavy
fields associated with
\(\mathrm{SU(3)_B}/\mathrm{SO(3)_B}\) can be integrated out, and the monopole is
seen at long distances as the generalized Cho--Maison monopole of the
\(\mathrm{SU(3)_A}\times\mathrm{SO(3)_B}\to\mathrm{SO(3)_{\rm diag}}\) theory.

This example illustrates the ingredients required for the UV completion.  One
needs an embedding \(G_{\rm B}\subset \widetilde G_{\rm B}\) with
\(\widetilde G_{\rm B}\) simply connected, a first-stage scalar whose VEV
breaks \(\widetilde G_{\rm B}\) to \(G_{\rm B}\) and gives a nontrivial
\(\pi_2(\widetilde G_{\rm B}/G_{\rm B})\), and a second-stage scalar whose
low-energy representation reproduces the Higgs field of the generalized
Cho--Maison model.  Therefore the mechanism is not restricted to the
electroweak example, although its realization is model dependent and requires
such an embedding of the low-energy gauge group and matter representations.

\section{Summary and Future Direction}\label{sec:summary}
The Cho--Maison monopole is an exotic monopole configuration in the Standard Model found by Cho and Maison \cite{Cho:1996qd}. In this paper, we have investigated two important aspects of the Cho--Maison monopole.

First, we clarified that the Cho--Maison monopole is not a special feature of the electroweak theory. As a concrete example, we explicitly constructed a generalized Cho--Maison monopole configuration in a toy model with symmetry breaking $\mathrm{SU(3)} \times \mathrm{SO(3)} \to \mathrm{SO(3)}_{\mathrm{diag}}$.
Moreover, these monopole configurations can be naturally generalized to models with a broad class of gauge symmetry breaking patterns, such as $\mathrm{SU(N)} \times \mathrm{U(1)} \to \mathrm{SU(N-1)} \times \mathrm{U(1)_{diag}}$ and $\mathrm{SU(N)} \times \mathrm{SO(N)} \to \mathrm{SO(N)_{diag}}$. 
As in the Wu--Yang monopole construction \cite{Wu:1976ge} and the original Cho--Maison monopole configuration \cite{Cho:1996qd}, a generalized Cho--Maison monopole configuration cannot be described by a single gauge patch surrounding the monopole, but instead requires two gauge patches that are glued together.

Second, we have shown that the Cho--Maison monopole can be interpreted as a low-energy effective description of an ’t~Hooft--Polyakov monopole in a UV theory. In other words, the Cho--Maison monopole can be ``UV-completed'' by an ’t~Hooft--Polyakov monopole. In this framework, the energy of the monopole becomes finite and is determined by the symmetry breaking scale in the UV theory.
In particular, we showed that an ’t~Hooft--Polyakov monopole in a model with two-step gauge symmetry breaking $\mathrm{SU(2)_A} \times \mathrm{SU(2)_B} \to \mathrm{SU(2)_A} \times \mathrm{U(1)_B} \to \mathrm{U(1)_{diag}}$ behaves as a Cho--Maison monopole in the long-distance region. Similarly, a monopole in the Pati--Salam model also behaves as the electroweak Cho--Maison monopole when observed at distances larger than the GUT scale. 

Furthermore, the same mechanism can also be applied to the generalized $\mathrm{SU(3)_A}\times\mathrm{SO(3)_B}$ monopole by embedding it into a two-step breaking of $\mathrm{SU(3)_A}\times\mathrm{SU(3)_B}$.  In such embeddings, the stability of the monopole is inherited from the underlying ’t~Hooft--Polyakov monopole, thereby providing a mathematical guarantee of its stability.

As a future direction, it is important to clarify the conditions under which Cho--Maison monopoles can arise. 
Although it is well known that the $\mathrm{SU(5)}$ grand unified theory admits ’t~Hooft--Polyakov monopoles, monopole solutions obtained from $\mathrm{SU(5)}$ do not exhibit Cho--Maison-type behavior because the Higgs doublet acquires a nonzero vacuum expectation value at the origin \cite{Eckert:1983ze, Eckert:1983bq}. 
This observation suggests that the realization of Cho--Maison monopoles would require specific conditions on both the gauge group and the symmetry breaking pattern. 
A systematic formulation or classification of such conditions
will be discussed in ref.~\cite{SatoShimamoriMiya}.

\section*{Acknowledgements}
The authors thank Soichiro Shimamori for useful discussions in the early stage of this work.
The authors also thank Takaya Iwai for reading our manuscript carefully and giving useful comments.
The work of RS is supported in part by JSPS KAKENHI Grant Numbers~23K03415, 24H02236, and 24H02244.
This work of FM is supported by JST SPRING, Grant Number JPMJSP2138.

\bibliographystyle{JHEP}
\bibliography{monopole}

\end{document}